\documentclass[12pt]{article}

\newcommand{\mathsym}[1]{{}}
\def\id{\protect{{1 \kern-.28em {\rm l}}}}

\def\be{\begin{eqnarray}}
\def\ee{\end{eqnarray}}

\makeatletter
\renewcommand\section{\@startsection {section}{1}{\z@}%
                                   {-3.5ex \@plus -1ex \@minus -.2ex}%
                                   {2.3ex \@plus.2ex}%
                                   {\normalfont\large\bfseries}}
\renewcommand\subsection{\@startsection{subsection}{2}{\z@}%
                                   {-3.25ex\@plus -1ex \@minus -.2ex}%
                                   {1.5ex \@plus .2ex}%
                                   {\normalfont\normalsize\bfseries}}

\makeatother

\def\b{{\rm b}} %$\beta$ in KRTT
\def\d{{\rm d}} %$-\alpha in KRTT

\def\Tr{{\rm Tr}}

\def \foot {\footnote}
\def \bi{\bibitem}

\def \ha {{1 \over 2}}
\def \td {\tilde}
\def \ci{\cite}
\def \N {{\mathcal N}}
\def \ww {\Omega}

\def\S{{\mathcal S} }

\def \E {{\mathcal  E}} \def \J {{\mathcal  J}}

\def \d {\del}

\def\a{\alpha}
\def\b{\beta}

\def\C{{\bf C}}
\def\P{{\bf P}}

\def \del{\partial}
\def \a {\alpha}

\def\g{\gamma}
\def\s{\sigma}

\def\ov{\over}

\def\J{{\mathcal J}}

\def\E{{\mathcal E}}

\def\b{\beta}
\def\l{\lambda}

\def \k {\kappa}
\def\foot{\footnote}

\def\det{\hbox{det}}
\def \ci {\cite}

\def \foot {\footnote}
\def \bi{\bibitem}
\def \ha {{1 \over 2}}

\def \fo { {1\ov 4}}

\def \d {\partial}

\def \el {\ell}
\def \Tr {{\rm Tr}}
\def \P {\Phi}
\def \l  {\lambda}

\def \N {{\mathcal N}}
\def \S {{\rm S}}

\def \td {\tilde}

\def \D {\Delta}
\def \N {{\mathcal N}}

\def \bi{\bibitem}
\def \la {\label}

\def \l {\lambda}
\def\foot{\footnote}

\def \sql {{\sqrt \l}}
\def \adss {$AdS_5 \times S^5~$ }
\newcommand{\rf}[1]{(\ref{#1})}
\def \ov {\over}

\def\N{{\cal N}}

\def\cc{\circ}

\def \ha{{1\ov 2}}

\def\r{{\rm r}}

\def \no {\nonumber}

\def \J {\mathcal{J}}
\def \del {\partial}

\def \E {{\cal E}}

\def \S {{\cal S}}
\def \J {{\cal J}}

 \def \bb {\bar \beta}

\def \bi{\bibitem}
\def \la {\label}

\def \l {\lambda}
\def\foot{\footnote}

\def \sql {{\sqrt \l}}

\def \adss {$AdS_5 \times S^5$\ }

\def \D {\Delta}

 \def \r {\rho}

\def \ov {\over}

\def \varpi {{\rm w}}
\def \OO {{\cal O}}

\def \pa{\partial}

\def \te {\theta}

\def \cc {{\rm f}}
\def \d {\delta}

\def \S  {{\rm S}}

\def \pa {\partial}
\def \C {{\cal C}}
\def \bea {\be}
\def \eea {\ee}
  \def \d {\delta}

\def\Tr{{\rm Tr}}

\def \b {\beta}

\def \del {\partial} 
\def \d {\partial}
\def \s {\sigma}

 \def \J {{\cal J}}
 \def \S {{\cal S}}
 \def \E {{\cal E}}

\def \d {\del}
\def \bd {\bar \del}
\def \ww  {\omega}

\def \os  {\OO({\textstyle{ 1\ov \sql}})}
\def \oss  {\OO({\textstyle{ 1\ov (\sql)^2}})}
\def \osss  {\OO({\textstyle{ 1\ov (\sql)^3}})}

\def \bd {\bar \del} \def \sql {\sqrt{\lambda}} 
\def \vp {\varphi}

\def \cC {{\cal C}}
\begin{document}

%%%%%%%%%%%%%%%%%%%%%%%%%%%%%%%%%%%%%%

\overfullrule=0pt
\parskip=2pt
\parindent=12pt
\headheight=0in \headsep=0in \topmargin=0in \oddsidemargin=0in

\vspace{ -3cm}
\thispagestyle{empty}
\vspace{-1cm}

\rightline{Imperial-TP-AT-2009-3}

\rightline{NSF-KITP-09-105}

\begin{center}
\vspace{1cm}
{\Large\bf  

%``Short'' 
Quantum  strings in $AdS_5\times S^5$: \\
 strong-coupling corrections  to  
dimension\\  of
  Konishi operator
%   in $N=4$ SYM   theory 

\vspace{1.2cm}

   }

\vspace{.2cm}
 {
R. Roiban$^{a,c,}$\footnote{radu@phys.psu.edu} and A.A. Tseytlin$^{b,c,}
$\footnote{Also at Lebedev  Institute, Moscow. tseytlin@imperial.ac.uk }}\\

\vskip 0.6cm

{\em 
$^{a}$Department of Physics, The Pennsylvania  State University,\\
University Park, PA 16802 , USA\\
\vskip 0.08cm
\vskip 0.08cm $^{b}$Blackett Laboratory, Imperial College,
London SW7 2AZ, U.K.\\
\vskip 0.08cm
\vskip 0.08cm 
$^{c}$Kavli Institute for Theoretical Physics,
University of California,\\
 Santa Barbara CA 93106,  USA
 }

\vspace{.2cm}

\end{center}

\begin{abstract}
 %%%%%%%%%%%%%%%%%%%%%%%%%%%%%%%%%
 We consider  leading strong coupling corrections to the energy 
 of the lightest massive string modes in \adss, which should be dual to members of the 
 Konishi operator multiplet  in ${\cal N}=4$  SYM theory. 
 %AT Through AdS/CFT  
 This determines the  general structure of the strong-coupling 
 expansion of the  anomalous   dimension of the  Konishi operator.  
 We  use  1-loop  results for  several  semiclassical string  states 
 to extract  information about the leading  coefficients in this
  expansion. Our prediction is 
  %AAT
  $ \D= 2 \lambda^{1/4}  + b_0   +  b_1 \lambda^{-1/4} +  b_3 \lambda^{-3/4}+ 
   ...$, 
  where $b_0$ and $b_1$  are  rational while $b_3$ is transcendental (containing $\zeta(3)$). 
  Explicitly, we argue that $b_0= \D_0-4$  (where 
  $\D_0$ is the canonical dimension of the corresponding gauge-theory  
  operator in  the Konishi multiplet)  and $b_1=1$. 
%vvv2
Our conclusions 
are sensitive to few assumptions, implied by a correspondence with 
  flat-space expressions, 
on how to translate 
% results found through
semiclassical quantization results  into predictions for the exact quantum string
spectrum.

%%%%%%%%%%%%%%%%%%%%%%%%%%%%%%%%%
\end{abstract}

\newpage
\setcounter{equation}{0} 
\setcounter{footnote}{0}
\setcounter{section}{0}

\renewcommand{\theequation}{1.\arabic{equation}}
 \setcounter{equation}{0}

\setcounter{equation}{0} \setcounter{footnote}{0}
\setcounter{section}{0}

\def \edd {\end{document}}

%\edd
\def \cc {{c }} 
\def \OO {{\cal O}}
\def \te {\textstyle}
\def \fl {\sqrt[4]{\l}}

\def \ha {{{\textstyle{1 \ov2}}}}
\def \fo {{\textstyle{1 \ov4}}}
\def \rx {{\rm x}}
\def \hg {{\hat \g}}

\def \C  {{\rm C}}
\def \hC  {{\rm \hat  C}}
\def \dd  {{\rm d}}
\def \bb {{\rm b}}
\def \dDelta {2}
\def \sql {{\sqrt{\l}}}

 \def \an {{\rm an}} \def \nan {{\rm nan}}
 
%%%%%%%%%%%%%%%%%%%%%%%%%%%%%%%%%%%%%%

\section{Introduction }
%%%%%%%%%%%%%%%%%%%%%%%%%%%%%%%%%%%
%refs to short ? Ambjorn   Janik  AF   Kazakov   ?
%To study the spectrum of anomalous dimensions we are to compute 2d dimensions 
%and impose marginality condition. 
%We may use standard NSR approach or l.c. gauge  or GS string approaches
%which all are available   in near-flat limit.
%if we look at small  massive string states  the zero-mode issues should not be relevant 
%and we may  compute their masses  using near flat space expansion.

The canonical example of the AdS/CFT duality \ci{mal,gkp1,witt}
implies the equivalence between  the spectrum of the planar 
$\N=4$  SYM theory  and  the spectrum of  free quantum string  in \adss  space. 
The spectrum  of the gauge   theory can be described either as a list of possible 
energies of SYM  states on $R \times S^3$  (as functions of  various quantum numbers)
 or 
as a list of dimensions $\D$  of conformal primary operators on $R^{1,3}$
(determined by   diagonalisation of anomalous dimension  matrix  for 
%, e.g., from singularities of the 2-point functions of 
single-trace gauge-invariant operators). 
Similarly, the string spectrum   is 
given by the  $AdS_5$  energies $E$  of string states on a cylinder $R \times S^1$ 
(found using,e.g.,  a light-cone gauge approach)
or is found  from the marginality condition for the corresponding 
string vertex operators on a plane   $R^{1,1}$
(by diagonalizing   of the  2-d anomalous dimension matrix).

Below we will be interested  in the strong coupling expansion of dimensions 
of  gauge theory operators  
or inverse string tension  expansion of energies of the corresponding 
 quantum  string states. 
%In general, one can compute energies on $R \times S^1$  in l.c. gauge \foot{This leads to massive
%integrable 
%theory and one may apply TBA to find its spectrum.}
%  (or in gauge theory on $R \times S^3$) 
%or dimensions of vertex operators on $R^2$ in conformal gauge  
% (dimensions of primary operators on $R^4$).
%Here we shall use first   approach in semiclasical approximation but mostly use second. 
%\foot{Size of $S^1$ is controlled by charges; in vertex operator approach charges 
%take   any fixed values.}

To set up the notation we will be using below, 
 we will  label representations of the bosonic subgroup 
  $SO(2,4) \times SO(6)$ of the symmetry group $PSU(2,2|4)$ 
   by  the Young tableaux   labels   \be 
   \la{yo} 
   \hC= (E, S_1, S_2; J_1, J_2, J_3)\equiv (E, \C)  \  . \ee
Here   each   charge of the highest-weight state   corresponds to six  $SO(2)$  subgroups
with $\C=(S_1, S_2; J_1, J_2, J_3)$ being the spins. 
These  are  related to the often used 
%(e.g., in \ci{bia}) 
 $SU(2)\times SU(2)$ labels $(s_L,s_R)$ for $SO(4)$ 
and the $SU(4)$ Dynkin labels $[p_1,q,p_2]$ as:
$s_{L,R} = \ha (S_1\pm S_2),$ and 
$p_{1,2}= J_2\mp J_3, \ q = J_1-J_2$, 
% , \ d_3 = J_2+J_3
 i.e.
\be \la{dyn}  \C= [ J_2-J_3,  J_1-J_2 , J_2+J_3]_{( {S_1+S_2\ov 2}, {S_1-S_2\ov 2})}   \ . \ee
Then  the equivalence  of the gauge and string theory spectra  can be  expressed as 
\be 
\D (\l,  \C) =   E (\sql, \C)   \ , \la{on} 
\ee
where $\D=  E_{\rm gauge},\ \ E= E_{\rm string}$, $\l$  is the  `t Hooft coupling and $ {\sql \ov 2 
\pi}= { R^2 \ov 2 \pi \a'}$ is the \adss 
string tension.\foot{Here we suppressed any potential
%R
 dependence of $\D$ or  $E$ 
on  various other ``hidden'' charges that specify the gauge theory   operators and the  quantum
string states.}  
 
 In the weak-coupling (${\l \ll 1}$) expansion  represented   by the perturbative gauge theory
 \be 
 \D
 %_{_{\l \ll 1}}
   = \D_0 + \g (\l, \C)  \ ,  
 \ \ \ \ \ \ \  \ \ \   \g  = k_1 \l + k_2 \l^2 + ...  \ ,     \la{tw} 
  \ee
   where $\D_0$ is the  canonical dimension of the corresponding operator.  
 $\g$ is an eigenvalue of the 4-d  anomalous dimension matrix.
 Only the operators with the same $ \D_0$ can mix, so $\D_0$ may be called a  ``level'' 
 of gauge-theory  states.  $ \D_0$   may change for states   within the same supermultiplet, as dictated by the commutation relations of $PSU(2,2|4)$, 
%R
while $\g$ should be the same. 
 
In the strong-coupling  (${\l \gg 1}$) 
expansion  represented   by  the perturbative (inverse string tension) expansion 
in   the string-theory  sigma  model, 
 one may expect   
that for large $\l$  (or large radius $R   \gg \sqrt{ \a'}$ of \adss  space)  
 massive quantum  string states   or ``short'' strings 
 with fixed charges $\C$  probe  a near-flat region of \adss 
 and thus  their energies  may  be  found by a 
 near-flat-space  expansion. 
 Then one may expect to find 
%  This leads to 
 \be \la{tri}
&& E(\sql, \C)
%_{_{\l \gg 1}}  
= 2\sqrt{n-1} \fl  \ +
% \  \b (\fl, \C) \ ,\ \ \ \ \ \ \ \ %\b =
 \ \sum_{k=0}^\infty { b_k \ov (\fl)^k} \ .
 %= \b_{\an} 
 %  + \b_{\nan} \ , \\
%\ \ \ \ \  \ \ \ \ \ && 
% \b_{\an} =   {b_1 \ov \fl }  +  {b_3 \ov (\fl)^3  } + \OO( {1 \ov (\fl)^5  } ) \ , 
% \ \ \ \ \ \ \ 
% \b_{\nan} = b_0 + {b_2 \ov (\fl)^2 }  + \OO( {1 \ov (\fl)^4  } ) \  . 
%\la{beb}
 \ee
 Here the leading term \ci{gkp1}  is the  analog of the flat-space string mass 
 term  (originating   from   $\a' E^2 = 4(n-1)$)  with 
 $n$ being the flat-space string level.\foot{
 %AAT
 We shall use the NSR definition of string level, with $n=1$ corresponding to massless level and $n=2$ to the first excited level.
 For
 some earlier discussions of  energies of quantum string states in \adss 
 see also 
 %v2
 \ci{pol,tsev,ft2,afs,afss,bl,tir}.
 In particular, an expansion of the form \rf{tri}  appeared  in the fermionic model 
 for the $su(1|1)$ sector in \ci{afss}.} 
 The structure of corrections   may be, 
  in principle,  determined  from 
  diagonalization of the 2-d anomalous 
 dimension matrix for the  corresponding 
 string vertex operators  (see  \ci{pol,tsev}  and  below) 
 having  the same canonical 2-d dimension $2n$, i.e. representing
 states  from  the same string 
  level $n$.\foot{The 2-d  operators that  may mix  must  have the same $n$. 
  This follows, e.g., 
     from momentum conservation when computing the 2-point functions  in 
   world-sheet perturbation theory.}
  %v2
   The 2-d anomalous  dimensions are given by a {\it regular }
   series expansion in $\a' = { 1 \ov \sql}$, 
   while $1 \ov (\fl)^k$  appear as a result of solving 
   quadratic-type  equations for $E$ following from the marginality condition.
   %vv2
   \foot{In particular,  there cannot be any $\log \l$  terms such as
   those that appear 
   in the strong-coupling expansion  of the anomalous dimensions 
   computed using  asymptotic Bethe ansatz equations \ci{brs}.}
    
    States    belonging to the same
  supermultiplet  must have the same $n$ but  
  may have different values of $b_0$,
  %R
    which should differ by the same amount as the canonical dimension $\D_0$ 
  in \rf{tw}.

 It is useful to  split the sum in \rf{tri} into  the ``odd'' (${b_1 \ov \fl }  +  {b_3 \ov (\fl)^3  } +...
  $)  and ``even''  ($b_0 + {b_2 \ov (\fl)^2 }  + ...$) power   parts
  as these  appear to  have   different origin
 within the semiclassical expansion we shall use  to determine
  the strong-coupling  coefficients $b_k$. Then one can also rewrite  \rf{tri} as 
 % ($b_k'\equiv  {b_k \ov 2\sqrt{n-1}}$)
   \be \la{trh}
&& E = E^{(\an)} + E^{(\nan)} \ ,\\  
&&E^{(\an)} = \sqrt{ \sql } \ \Big[ 2\sqrt{n-1}   +  {b_1 \ov \sql  }  +  {b_3 \ov (\sql)^2  } +
 %\OO( {1 \ov (\sql)^3  } ) 
 \osss \Big]  \ ,
 % \ \ \ \ \ \ \ b_k'\equiv  {b_k \ov 2\sqrt{n-1}} 
  \la{non}   \\
&&E^{(\nan)}  = b_0 + {b_2 \ov \sql  }  + \oss
%\OO( {1 \ov (\sql)^2  } ) 
\ . \la{aan}
\ee
 As we shall see, in the semiclassical approach to the energy of strings with small values of spins  the ``analytic'' part   $E^{(\an)}$   is the one that
 originates  from the classical string energy  and  also from a 
 ``regular'' part of semiclassical corrections (e.g., determined by even powers of masses 
 of string fluctuations)  while the ``non-analytic'' part $  E^{(\nan)}$  has its 
 origin, from semiclassical standpoint, 
  in certain special  IR  parts of quantum corrections (which are due to zero or ``light'' modes
 that become massless in the ``small-spin'' string limit).
 %v2
 \foot{This distinction into ``analytic'' and ``non-analytic'' 
 terms in the 1-loop energy in the small-spin limit, which has an IR
origin,  should not be confused  with the one in \ci{bets} which
appeared in the large-spin limit  and had an UV origin.}
 
 % Our non-trivial prediction is  that $\b$ in  \rf{tri} contains only  odd
%  inverse powers of $\fl$. 

The weak-coupling expansion  \rf{tw} given by  the planar 4-d   perturbation theory  
should have  finite radius of convergence and thus 
should define $ \D(\l, \C)$ for all values of $\l$. 
Expanding the resulting function at large $\l$ one should then reproduce 
the strong-coupling   expansion \rf{tri}  as predicted by the string theory. 
Once $\l$ is  increased so that the anomalous dimension $\g$  becomes of the same 
order as $\D_0$,
%R 
the latter   looses its ``invariant'' meaning. 
An interesting question is 
%It is not clear a priori 
how the value of $\D_0$ is encoded   in the strong-coupling 
expansion  coefficients in  \rf{tri}.  % Vice versa, 
And vice  versa,  the  meaning  of  the string level $n$  in 
\rf{tri}  in the weak-coupling gauge theory expansion \rf{tw}
 is also  unclear a priori. 
 
 \
 
 Our aim below   will be to clarify 
% Below we shall   determine
  the  general structure of the strong coupling   expansion \rf{tri}, \rf{trh} 
 on examples of string states at the first excited level $n=2$ which are  dual to 
% some 
 %of the  dimensions of the 
 members of the  Konishi operator  multiplet in gauge theory.  
 %  which should be dual to 
 % string states at the first excited level $n=1$. 
  % justifying  the expansion \rf{tri}.
 We shall    use  1-loop string  results for  several  semiclassical string  states 
 to extract  information about the two leading  coefficients $b_0$ and $b_1$  in 
 \rf{non}, \rf{aan}.  Our results for the two subleading coefficients 
 in the  dimension  of the members of the Konishi operator
 multiplet (with $\D_0=2,..., 10$)  may be summarized as follows:
  \be \la{su}
 n=2: \ \ \ \ \ \   \     b_0 = \D_0 -4 \ , \ \ \  \ \ \ \ \ \   b_1 = 1   \ . \ee
 We shall also conjecture that 
  \be \la{cog}    b_2 =0  \ . \ee 
 The rest of the paper  is organized as follows. 
 We shall start in section 2 
 with general remarks on the structure of the strong-coupling expansion \rf{tri}
 explaining how  it follows  from  solving  the marginality conditions for the 
 corresponding string vertex operators. We shall  consider constraints 
 on the  1-loop  2-d anomalous dimension   implied  by  the structure of 
 the Konishi supermultiplet 
 %R
 and identify, from this standpoint, the origin of the two components 
 $E^{\rm (an)}$ and $E^{\rm (nan)}$. We shall also discuss  form of  the 2-d 
 anomalous dimensions of the corresponding composite  operators 
 as determined  by  the  bosonic part of the \adss  string sigma model 
 
 To systematically include the effects of fermions  in section 3 we propose  to use 
 a  different strategy: start with semiclassical spinning 
 string solutions, compute 1-loop corrections to their energies and then attempt to interpolate 
 to small values of spins corresponding to states at the  first excited string level. 
 We end up with what appears to be a consistent picture with different types of 
 spinning  string states  finding their counterparts  among the states in the Konishi multiplet table 
 and predicting the same universal  expression for the corresponding anomalous dimension. 
 Our results are summarised in section 4.

\renewcommand{\theequation}{2.\arabic{equation}}
 \setcounter{equation}{0}

 \section{General structure of strong-coupling expansion}

 There are few guiding principles that one may try to use to 
 understand the interpolation of dimensions of composite operators from weak to strong coupling. 
 First, one   may expect the  validity of a 
 ``non-intersection principle'' \ci{pol}: 
there should   be  no level crossings for states  with the same  quantum numbers 
as  $\l$  changes from weak  to strong  coupling. 
That would 
suggest   that (for fixed values of  charges)  the states
with smaller  values of the gauge-theory ``level''  $\D_0$  and thus smaller  dimension
at weak coupling  should correspond to states
 with  smaller   energy also at strong coupling. 
 %values of string level  $n$.
  The singlet  Konishi scalar operator  with $\D_0=2$  which has lowest 
 dimension  at weak coupling   should correspond to 
 a string state on the first excited level $n=2$. In fact, 
 the analysis based on  symmetries and near flat space  expansion 
  suggests that 
  %v2
  the states of the Konishi supermultiplet \ci{ferr} 
  should belong  \ci{bia,bia2}  to the set of 
   the superstring states 
 at the  level $n=2$. \foot{
 More precisely, the Konishi supermultiplet should be the $J=0$  Kaluza-Klein ``floor''
 of the whole set of states at the first excited level given by 
 $ \sum_{J=0}^\infty [0,J,0] \times $ [Konishi multiplet] \ 
 \ci{bia}. Adding extra $S^5$ orbital momentum $J$ increases  canonical dimension of the 
  gauge-theory operator but does not increase the level of the 
  dual string state.}

Second, since gauge-theory states  belonging  to the same supermultiplet should have the same 
anomalous dimension (while  their $\D_0$'s  may  differ by  (half)integer values 
as they are related   by application of  supersymmetry generators) the equality \rf{on} 
suggests, in view of \rf{tri},  that for 
 the corresponding string states
\be \la{fow}
E
%_{_{\l \gg 1}}   
=  2\sqrt{n-1} \ \fl   \  + \D_0   + \bb_0  +  {b_1 \ov \fl } + {b_2 \ov \sql  } + 
\OO( {1 \ov (\fl)^3  } ) 
% {b_2 \ov (\fl)^3  } +... 
  \ , \ \ \ \
  % {\rm i.e.} 
  \ \ \ \
b_0 =   \D_0 + \bb_0\ ,  \ee
where   the coefficients    $n, \bb_0,  b_1, b_2, ...$  appearing in 
the strong-coupling expansion of the anomalous dimension 
$\g$ should be universal, i.e. should be  the same   for all the states in a supermultiplet.\foot{ 
This expression is indeed consistent with the non-intersection principle: 
the states with the same  charges (at the same  string level or in the same supermultiplet) 
that had  smaller   dimension (i.e. smaller  $\D_0$) at weak coupling  will have smaller 
dimension also at strong coupling.}

In contrast to the weak-coupling region \rf{tw} where members of the same supermultiplet may have 
very different dimensions as $\D_0$ may  jump from state to state, at strong coupling \rf{tri}
all dimensions of states from the same level  are  approximately equal, differing only 
in the subleading terms 
controlled again by $\D_0$  part of $b_0$. 
While in flat space all string states at a given level have
the same  mass or rest-frame energy, switching on the curvature  
removes this degeneracy.\foot{Note also that,
%R
similarly to weakly-coupled gauge theory
%  as in the  gauge theory at weak coupling  
where the  operators  can be constructed in terms of the free-theory fields, 
at strong coupling or in the near-flat-space expansion 
one may  label string states by the oscillator numbers of the flat-space superstring description.}

\

%Below we will be interested  in  finding  the values  of the leading strong coupling coefficients 
%$\bb_0$ and   $b_1$ for the first excited level ($n=2$) string states  that  should   belong 
%to the  Konishi multiplet.

The  main  problem is how to compute the quantum dimensions of the corresponding vertex 
operators and thus determine the coefficients in the expansion \rf{tri} 
of $E$. 
As we shall discuss in more detail in section 2.2 below, one  expects 
the leading terms  in the (eigenvalue of the)  2-d 
dimension of the  vertex operator  representing string states with   charges 
$\hC= (E, \C)$  to be a generalization of the 
flat-space marginality condition 
%(written in the rest frame $p_i=0$):
$2 =  2 n  - { \a' \ov 2} (E^2 - p_i^2) $. 
In \adss  the term    $E^2$  is  replaced  by a certain quadratic combination 
of the relevant charges.  For example, for a state  carrying a spin $J$ 
we may a priori  expect 
% (for simplicity here we treat $\C$ as a single charge)
\be \la{dim}
2 =  2 n  - { 1 \ov 2\sql } \Big[E (E + a_1)    + a_2 E J   + a_3 J( J  + a_4)   + a_5 \Big]
+ \oss
%\OO({1 \ov (\sql)^2  })
 \ . 
\ee
The  expansion of the 2-d anomalous  dimension   goes in 
integer powers of the inverse string tension, i.e. 
contains only $ {1 \ov (\sql)^k  }$-terms.\foot{
To compute  $1 \ov \sql$ corrections  to the canonical 2-d dimension $2n$-term  
one is supposed to choose a 
 basis of composite operators consistent with symmetries,   compute
 the  anomalous dimension matrix   using string sigma model perturbation theory  and then  
diagonalize this  matrix.  The resulting eigen-operators  will   be given by linear 
 combinations of operators  from the basis 
 (with coefficients that may  depend on  ${1 \ov \sql}$). These will be  
 conformal primaries that have definite dimensions and define 
string vertex operators (that can be used also to compute 
correlation functions and thus string  scattering amplitudes, etc.).}

The  structure of \rf{dim}   is implied also by   the space-time  interpretation 
of the 2-d  anomalous dimension operator as a differential operator acting 
on the corresponding tensor  coefficients  $\Psi$  of a basis of vertex operators.
%R
For a flat-space state with mass $m_0^2 = {4(n-1) \ov \a'} $, one expects to 
find in curved background
%
%In curved  background one expects to find ($m_0^2 = {4(n-1) \ov \a'} $) 
\be \la{cur}
\Big[ 2 - 2n     + { \a' \ov 2} \nabla^2  +  \a' (c_1 {\rm R} +   c_2 {\rm F_5 F_5})  + \OO(\a'^2 ) \Big] \Psi
=0 \ , \ee
where R stands for   the curvature tensor  and  F$_5$ stands for the 5-form field strength
(the $\a'$ term may contain several tensor structures, cf. \ci{bl}). 
Due to the large amount of supersymmetry, higher $\a'^k$   corrections  to the ``mass matrix'' are 
expected (on the basis of the NS-NS sector experience) to start at 
 relatively late order.\foot{This  suggests  that non-trivial  corrections in 
\rf{dim} should be postponed  at least till 
%start probably at 
order  ${1 \ov (\sql)^3}$.}

The expression \rf{fow}  for $E(\sql, \C)$  then follows from 
\rf{dim}  by solving it perturbatively in $1 \ov \sql$. 
For the lowest level (supergravity or BPS)  states with $n=1$ 
each ${1 \ov (\sql)^n}$ term  in \rf{dim}  should  vanish separately
so that $E$  should  not depend on $\sql$.
For massive string states   with $n > 1$   and for fixed charges $\C=(J,...)$ 
%and ignoring higher order corrections 
we get from \rf{dim}
\bea  \la{eex}
E^2 - 2  q_0 E + q_1   =  4 (n-1) \sql  + \os
%\OO( {1 \ov \sql}) 
 \ . \eea
% \\
Solving this quadratic equation produces terms with powers of 
square root of string tension, 
i.e.   leads to  \rf{fow}  with 
%&& E= 2\sqrt{ n-1}\  \fl  + b_0   +  { b_1 \ov \fl } + \OO({1 \ov (\fl)^3  }) \ ,\ \ \ \ \ \ \ \ \ 
\be  b_0= q_0  \ , \ \ \ \ \  \ \ \ \ \ 
b_1= { q_0^2 - q_1 \ov 4 \sqrt{ n-1}} \la{hhl} \ . 
\ee
i.e. reproduces the  structure of the strong-coupling expansion anticipated in   \rf{tri}.
It is then clear that  any effective $ {1 \ov \sql}$  corrections to the $q_1$ and $q_2$ 
terms coming from   $\OO( {1 \ov \sql})$ term in  \rf{eex}  
will be subleading compared to the three leading terms in \rf{hhl}.
  Thus $b_0$ and $b_1$ are determined by the 1-loop correction to the 
  2-d anomalous dimension in \rf{dim}.

Let us note also that  the $b_2$-term in \rf{aan}  may appear only from the
{\it 2-loop}    ${1 \ov (\sql)^2 } E$ term  in \rf{dim} (which effectively shifts 
the coefficient $q_0 \to q_0 +  {c \ov \sql}$ in \rf{eex}). 
As was mentioned above,  it seems  
 likely that such terms should  not appear    due to supersymmetry so we  conjecture \rf{cog}  that 
 $  b_2 =0$.

\subsection{Supersymmetry constraints:  Konishi supermultiplet}

In the case of the states  from the first excited string level 
that are expected to correspond to states
of the Konishi multiplet  we  find from \rf{fow}, \rf{eex}
\bea 
&&
E^2 - 2  (\D_0   + \bb_0)   E +  (\D_0 + \bb_0)^2 - 4 b_1     =  4  \sql   + \os
%\OO( {1 \ov \sql})
  \ ,
\la{jk}\\
&&
E  =  2 \fl   \  + \D_0   + \bb_0\    +  {b_1 \ov \fl } \ +  \OO( {1 \ov (\fl)^2  } )  \ .\la{ch} 
 \eea
Here $\bb_0$ and $b_1$   should be universal within the multiplet while 
$\D_0$ may change from 2 to 10   in steps of 1/2. 

%Our aim below will be to try to determine the values of $\bb_0$  and $b_1$. 
To  further clarify  the origin of  \rf{jk}, \rf{ch} let us 
study  to which  degree  the  strong-coupling  expansion of  $E$ is controlled by 
$PSU(2,2|4)$  symmetry that determines the structure of the Konishi multiplet
\ci{ferr,bia}  listed in Table  \ref{Ktable} (we borrow this table from \ci{bia}).
For every state  in the  Table \ref{Ktable}
there should exist a vertex operator at the level $n=2$. For  
%(first massive level)
each  operator we should get 
 the same value for the 4-d anomalous dimension  $\g= \D - \D_0= E- \D_0  $ 
(which is the only quantity undetermined by the representation theory) 
by solving  the 2-d  marginality condition. 

%The correction to the classical dimension of the vertex operators
%may depend on charges as well as on the level above the supermultiplet
%``vacuum'' at which the operator appears. This is so because, besides the
%quantum corrections inherited from the supermultiplet
%``vacuum'', there may exist additional renormalization due to both the
%supersymmetry generator and the lowest operator sitting at the same point. 

As discussed below in section 2.2, the 
 1-loop correction to the 2-d anomalous dimension in \rf{dim} 
 may be at most  quadratic in  the charges $\hC_A= (E, S_1, S_2; J_1, J_2, J_3)$
 (the corresponding  1-loop Feynman diagrams involve at most  two  of the fields of the
operator at a time). Then the general form of the 1-loop  marginality  condition 
will be 
(cf. \rf{dim} for $n=2$) 
%one can make the following ansatz for the anomalous dimension 
%of a generic vertex operator corresponding to a state
%$|\phi_n\rangle=Q^n|\phi_0\rangle$ at level 2:
\be \la{cb}
2=4  -  \frac{1}{2\sqrt{\lambda}}\Big(
\sum_{A,B=1}^6 u_{AB} \hC_A \hC_B +\sum_{A=1}^6 v_{\ell A} \hC_A +  h_\ell\Big)  + \oss
% \OO( { 1 \ov (\sql)^2})
  \ . 
%\equiv (2-4)+\frac{1}{2\sqrt{\lambda}} \gamma^{(1)}
\ee
Here $u,v,h$   are constant coefficients and we introduced 
dependence on the  supersymmetry ``level'' $\ell=0,1, .., 16$ of the supermultiplet, 
with 
\be  \la{dd} \D_0= 2 + \ha \ell    \ . \ee 
One may argue that the coefficients $  u_{AB} $ should  not
depend on $\ell$: 
the action of the supersymmetry generators only changes  charges of a given state 
by a finite amount (e.g.,  $J_1\rightarrow
J_1+ \ha \ell$, etc)  so  that the terms quadratic in the charges
do not acquire any $\ell$-dependence.
% while the terms linear in charges or
%independent of charges do pick up such a dependence.

Solving the  condition \rf{cb} 
 for   $E$  for all of the  bosonic states in
the Konishi multiplet whose charges are listed in Table  \ref{Ktable} \foot{It suffices to do  this  only for states up to (and
including) those with $\Delta_0 =6$ since the states with higher $\D_0$ 
can be found  by conjugation.} 
and requiring that $E$ jumps by $\ha (\ell_2 - \ell_1) $ when  going from  a
 supermultiplet level $\ell_2$  to  level $\ell_1$  one can determine  the coefficients 
 in \rf{cb}   and finally obtain 
the following expression for $E$ in \rf{ch}
with 
\be \la{vv}
\bb_0 = -2 + \frac{1}{2}(h_{2}-h_{0}-1) \ , \ \ \ \ \ \ \ \ \ 
b_1 =  \frac{1}{16}(h_{2}-h_{0}-1)^2- \frac{1}{4}  h_0   \ . 
\ee 
Here  $h_{0}$ and $h_{2}$ are undetermined universal (i.e. $\ell$-independent) constants.
The   marginality conditions for states at different supermultiplet levels $\ell 
= 2 \D_0 - 4 $    then follow from \rf{ch}. 
%up to ${\cal O}(\lambda^{-1/2})$:
%\be
%%%%%%%%%%%%%%%
These give expressions for the 2-d anomalous dimensions of the vertex operators
obtained by acting with $\ell $ supersymmetry generators on the one corresponding to the ``lowest''
state in the supermultiplet.\foot{Repeating similar analysis 
 in the case of the  short  multiplet  of  BPS (supergravity) states
starting with the  $[0,J,0]_{(0,0)}$   KK scalar state  one finds that the analog of \rf{dim} is 
$$2 = 2  - { 1 \ov 2 \sql} \Big[ E(E-4)  - ( J+ \ell) (J+4 - \ell) \Big] + \oss
 %\OO( { 1 \ov (\sql)^2})
 \ , $$
  where $\ell=1,2,3,4$ corresponds to the ``level''
 of the bosonic states in the supermultiplet.}

%AT
It is interesting to note that for the values of $b_0$ and $b_1$  in \rf{su}  we shall 
find below (i.e. $\bb_0=-4, \ b_1=1$) the relations \rf{vv} imply 
\be \la{hah}
h_0 =0 \ , \ \ \ \ \ \ \ h_2=-3   \ . \ee
The value  of $h_0 =0$ in \rf{cb} 
 appears indeed to be very natural for lowest-level state
in the Konishi supermultiplet.

%lowest one
%can be combined in a single acuum.

\subsection{Structure of 2-d anomalous  dimensions  of  vertex  operators}

To give an idea of   how one could   compute the 
2-d anomalous dimension  (and thus the 
values of $\bb_0$ and $b_1$ in \rf{ch}) 
from  first principles   let us review the structure  of the corresponding (bosonic) vertex 
operators following \ci{pol,tsev}.
The action of the  \adss 
superstring sigma model   \ci{mt} written in terms of the 6+6 embedding coordinates  has 
the following structure 
% [Metsaev, AT 98]
\bea 
\la{scc} && I= { \sql \ov 4 \pi}  \int d^2 \s  \ \Big(  -\del N_a \bd N^a +  \del n_k \bd n_k 
+ {\rm fermions\ } \Big)  \ , \\
&&  N_a  N^a = N_+N_+^* - N_x N_x^*  - N_y N_y^*  =1  , \ \ \ \ \ \ 
 n_k n_k = n_x n_x^*  + n_y n_y^* + n_z n_z^*=1,   \la{nen} \eea
where  
$  N_+ = N_0 + i N_5, \   N_x= N_1 + i N_2, \  N_y= N_3 + i N_4,$ \  $
n_x=n_1 + i n_2, \  n_x=n_3 + i n_4, \ n_z=n_5 + i n_6.$ 
The fermions make this model UV finite. 
The  aim is to construct  marginal (1,1)  vertex   operators  in terms of 
$N_a$,  $n_k$ and the fermions
which correspond  to the highest weight states  of $SO(2,4) \times SO(6)$ representations. 

For example,  the   vertex operator for dilaton-type 
massless level $n=1$  (supergravity)  scalar  mode with $SO(6)$ spin $J$ 
 should have the structure\foot{Recall that  $N_+ = \cosh \rho \ e^{i t}$ 
 where $t$ is  $AdS_5$ global time coordinate   and also 
  $n_x  \sim  e^{i  \vp }$   where $\vp$ is an isometric angle of $S^5$.}
\be \la{vew}
 V^{(0)}_J    =   (N_+)^{- E}\  (n_x)^J  \ (  -\del N_a \bar \del N^a  + \del n_k  \bd n_k  + 
   {\rm fermions})
    \ . \ee
 The  corresponding marginality condition is (cf. \rf{dim}) 
  \be 
   2= 2   - { 1 \ov 2\sql} \Big[ E(E-4) - J(J+4) \Big]  + \oss  \ , \ee 
so that to the 1-loop order  
 $E=  4 + J  $   and all  higher-order corrections should vanish as this should be a 
BPS state. 

In flat-space  string theory  a spin $S$   state  on the  leading Regge trajectory 
%with spin  $S$   and energy $E$ 
is represented by (ignoring fermionic terms)
$
V_S= e^{ -i E t } \big( \pa \rx_x\bar{\pa} \rx_x \big)^{S\ov 2}$, \ 
$ \rx_x = x_1+ix_2$, with the marginality  condition   being 
$ 2=S  - \ha \a'   E^2 =0,$  i.e. $E = \sqrt{{2\ov \a'} (S-2)} $.
By analogy, in \adss  case some  
  candidate operators for states on the leading Regge trajectory are 
 \be  
   V_J  =  (N_+)^{-E}\big( \pa n_x \bar{\pa} n_x \big)^{J\ov 2} +...\ , 
\ \ \ \ \ \ \ \ 
V_S =  (N_+)^{-E}\big( \pa N_x \bar{\pa} N_x \big)^{S\ov 2}+...\ , \la{mml}
\ee
where dots stand for the    fermionic terms 
and  $\a'\sim {1 \ov \sql} $ terms resulting 
 from  diagonalization of the anomalous  dimension  operator. 
 In general, ignoring the fermions,   the  operator 
$ \big( \pa n_x \bar{\pa} n_x \big)^{J\ov 2}$  in  the $SO(6)$ sigma model 
may mix  with 
   \be   (n_x)^{2p + 2q  } (\del n_x)^{{J\ov 2 } - 2p}   (\bar
\del n_x)^{{J\ov 2 }  - 2q }
( \del n_m \del n_m)^p ( \bd n_k \del n_k)^q \ , \la{cn}   \ee
where 
 $ p,q= 0,..., {J\ov 4} ; \    m,k=1,...,6$.
The  operator 
$(N_+)^{- E}\big( \pa N_x \bar{\pa} N_x \big)^{S\ov 2}$  in the  $SO(2,4)$  sigma 
 model  may mix  with  
\be \la{cha}
  (N_+)^{- E - p-q} N_x^{p+q} (\d N_+)^p (\d N_x)^{{S\ov 2}-p} 
(\bd N_+)^q (\bd N_x)^{{S\ov 2}-q} +     O(  \del N_a \del N^a \bd N_b \bd N^b)  \ , 
\ee
  where 
  $     p,q= 0,..., {S\ov 4}  ; \   a,b=0,1,...5$.
  The   true vertex operators are   
  eigenstates of the  anomalous dimension matrix, i.e.   
   particular linear combinations of the above structures. 
   
 These could, in principle, be found by solving Lichnerowitz-operator  type 
  equations  expressing  marginality condition.  In the 
  case of the bosonic model 
   $I = { 1 \ov 4\pi \a' } \int d^2 \s \  G_{mn}(x) 
\del x^m \bd x^n $    perturbed, e.g.,  by 
  $V = \Psi_{m_1 ...m_J} (x) \del x^{m_1} ... {\bd} x^{m_J} $
one could find the 2-d anomalous dimension  by 
computing  the renormalization of  $\Psi_{m_1 ...m_J}$ 
 and setting  $\beta_\Psi= \hat   \gamma  \Psi + \OO(\Psi^2)$=0.
 That would  give (cf. \ci{fridcal})
\be  
\hat \gamma  \Psi = \Big[2- J +  \ha \a'  \nabla^2   + 
  \sum c_k \a'^k (R....)^n...\nabla^p   \Big] \Psi =0  \ . \la{cury} \ee 
Solving  this equation for $\Psi$  would  amount  to finding the eigen-states
of $\hat \gamma$. 
%diagonalizing the  ``anomalous dimension'' operator
%e.g.  Lichnerowicz-type  equation for massive spin 2 
However, the  general    form of $\hat \g$ 
  for generic  $\Psi$ and  curved background   is not known  even 
  to  the leading (1-loop) order in $\a'$.\foot{Few exceptions are  the 
   WZW models  (and  models   related to them by  simple transformations like $T$-duality)
and some  plane-wave  models.}
 For that reason one apparently is to resort to  
     ``first-principles''  computation  for each  specific model.
  
 % [ how integrability of string theory may be reflected in the structure of 
  %$\hat \g$ ? ]
%thinking of an integrable   string theory   for which anom dim operator should have some  %special features due to this integrability --   how they would be reflected in anom dim %operator as differential operator in target space? In gauge theory we say -- dilatation %operator is a spin chain Hamiltonian, i.e. one of the charges of an integrable spin chain... %but here we deal with some background dependent second (and higher) order tensor  diff operator %in target space...
 
% use  global coordinates with linearly realized symmetry:
%  e.g.  for   $S^5= SO(6)/SO(5)$
% $$
% S ={  \sql  \ov  \pi  } \int d^2 \xi\  \del n_m \bd n_m   \ , \ \ \ \ \ 
%n_m n_m =1  \ $$ 
% $$
% \dot \ge   = - \epsilon  \ge 
%+ 4  \ge^2  +  4  \ge^3  + ... \ ,\ \ \    \ \ \ \ge \equiv  {1 \ov  \sql } = { \a' \ov \L^2} \ , \ 
%\ \ \ \ \   \epsilon = d-2   $$
% running is cancelled if embedded into \adss string theory 
 For example,  the operators  in the $SO(6)$ model that are relevant 
for states on  leading Regge trajectory (i.e. containing no terms with $\del ^k n $, $k > 1$) 
 % multiple derivatives on coordinates)we are interested  in  renormalization of  
 are \be 
O_{\el,s} =  \Psi_{k_1...k_{\el} m_1...m_{2s}}  
n_{k_1} ...  n_{k_\el}  \d n_{m_1} \bd n_{m_2} ...
\d n_{m_{2s-1}} \bd n_{m_{2s}}
  \ . \la{iop} \ee
Their renormalization was studied  in \ci{krav,weg,cast,tsev}. 
The 
simplest case is 
$ \Psi_{k_1 ...k_\el} n_{k_1} ... n_{k_\el}$  with 
traceless $ \Psi_{k_1 ...k_\el}$ which is    mapped by renormalization into 
itself and 
has the  same 2-d  anomalous  dimension   as its 
highest-weight representative 
$(n_x)^J $, i.e.   $ -    { 1 \ov 2 \sql}  J(J + 4)    + \oss  $;
it  corresponds to a scalar  spherical harmonic that solves  the  Laplace 
equation  on $S^5$.

Similar results are found for 
$SO(2,4)$ model by 
replacing $(n_x)^J$ and $\d n_k \bd n_k $   with, respectively,   
 $(N_+)^{-E}$ and $\del  N_a \bd N^a $,  
 and reversing the sign of the coupling,  $ { 1 \ov \sql} \to - { 1 \ov \sql}$.
Then  the  dimension of  $(n_x)^J \d n_k \bd n_k$, i.e.  
$ -2 -    { 1 \ov 2 \sql}     J(J + 4)   + \oss $
translates into  the dimension of 
 $(N_+)^{-E} \d N_a \bd N^a $, i.e. 
$-2   +  { 1 \ov 2 \sql}    E(E - 4)   +    \oss$, etc. 

 The number of $\d n_k \bd  n_k$  factors in an operator like 
 \rf{iop} never increases \ci{weg}   and thus 
can be used as a ``quantum number'' 
to characterise the  leading term in an eigen-operator.
 An example of a scalar operator carrying no  spins 
 is 
% example of scalar  higher-level   operator:  
 \be   V_r= (N_+)^{- E} \Big[(\d n_k \bd n_k)^r+ ... \Big] \ , \la{kop} \ee
 for which  the 1-loop and 2-loop terms in the 2-d  dimension  in 
 bosonic \adss model are \ci{krav,weg,cast,pol}
 \be && \hg(V_r) =  2- 2r   +   { 1 \ov 2\sql} \Big[   E (E-4) +  2r(r-1)\Big]  \cr
&& \ \ \ \  \ \ \ \  \ \  \ \ \ +  { 1 \ov (\sql)^2 } \Big[{\te{ 2 \ov 3}} r (r-1) (r- {\te{7 \ov 2}})  +
 4r \Big]  + \osss
%\OO({1 \ov (\sql)^3}) 
\ .  \la{klp}
\ee
This  operator  corresponds to a scalar string state at level $n=r$, 
so  the fermionic contributions   should make the  $r=1$ state  BPS, 
with $E=4$  following from the $\hg=0$ condition. 
The $r=2$ choice  should correspond to a scalar state 
on the first excited string level. Eq.\rf{klp}  implies then
(cf. \rf{eex}, \rf{jk}):
$E(E-4) = 4 \sql - 4  + \os $, so that 
$E= 2 \fl  + 2 + { 0 \ov \fl} +  \OO({1 \ov (\fl)^3})$.
This  result should  not, however, be trusted as the fermions are expected to change the $E$-independent 
terms in the 1-loop anomalous dimension. 
 
% still   for a scalar operator expect no
% leading correction to 
% $\hat \g = -\ha D^2 $ 
%  fermionic contribution should cancel 1-loop  mass shift $r(r-1)$?!  
%if  that happens  
%$$ \D(\D-4) = 4 \sql   + \os \ , \ \ \ \ 
%\D= 2 + 2 \sqrt{ \sql} \  [1    +  {1 \ov 2 \sql } +   \oss ]
%$$

An example of another singlet   scalar operator is 
$
 (N_+)^{- E}  (\d n_k \d n_k \bd n_m \bd n_m)^{q} $ with 
$   \hg= 2- 4q  +   { 1 \ov 2\sql} \Big[ E (E-4) +  16q  \Big]       +  \oss $, 
with   $q=1$  corresponding to  a state on the   first excited string level.  

Going back to the  operator in \rf{mml}  for a string state  with a 
spin  $J$  in $S^5$, we get 
\bea  \la{spii}
% && V_J=    (N_+)^{- E} \Big[ ( \d n_x   \bd n_x)^{J/2} + ...  \Big] \ , \\ 
 && \hg (V_J)   = 2-J     + 
{ 1 \ov 2 \sqrt \l} \Big[   E (E-4)  -  \ha  J  (J + 10 )  \Big] +  \oss  \ . \la{pii}
\eea
The 
inclusion of the  fermionic contributions 
  may   shift  the coefficient of the term linear in $J$. 
%$ J  (J + 10 ) \to J(J-2) $ 

An example   of a (bosonic)  operator  with 
two spins $(J_1,J_2)$ in $S^5$ is \ci{weg}  (cf. \rf{nen}) 
\be V_{J_1,J_2} =  (N_+)^{-E} 
\sum_{u,v=0}^{J_2/2}  c_{uv}   n_y^{J_1-u-v} n_x^{u+v} (\d n_y)^u (\d n_x)^{{J_2\ov 2}-u} 
   (\bd n_y)^v (\bd n_x)^{{J_2\ov 2} -v  }  \ ,    \la{kol}
 \ee  
 where $c_{uv}$  are constant coefficients. 
Ignoring  the  fermionic contributions,  the 
highest and the lowest  eigenvalues of the  resulting 
1-loop anomalous  dimension  matrix  are \ci{tsev} 
\be &&
\hg_{\rm min} =  2- J_2 
  +  { 1 \ov 2 \sql}  \Big[   E ( E - 4)  -  J_1(J_1+4) -    2  J_1 J_2 -   \ha   J_2 ( J_2 + 10  )    
     \Big]  
  +  \oss,   \no \\
&&  
\hg_{\rm max} 
 =   2-  J_2 
  +  { 1 \ov 2 \sql}  \Big[  E ( E - 4)   -  J_1(J_1+4)   -  J_2(J_2+6)    \Big] + \oss  \la{maa} 
  \ .    \ee
 The fermionic contributions may again alter the  coefficients of the terms linear in $J_i$
 and may be also  produce a constant term like   $h_\ell$ in \rf{cb}.
 
 Unfortunately, we do not know at present  how to systematically 
 incorporate  the fermionic terms into the  above vertex operators 
 and thus how to compute  the fermionic contributions to the  2-d anomalous dimensions
 starting with the \adss superstring action of \ci{mt}.
 %pure spinor approach a la Mikhailov- Schafer-Nameki may have some  use ?
 
%One  potential approach to determining the structure of the  1-loop anomalous dimension 
% is  to find the supersymmetry constraints  on the coefficients of the quadratic form 
% in \rf{cb} and then use semiclassical limit
%of the resulting expression to fix some of the remaining coefficients.

 One   possible indirect approach towards determining  these anomalous dimensions 
 may be to reconstruct  the quadratic  term in the space-time effective action 
 for the coefficient functions $\Psi$ in, e.g.,  \rf{iop} and thus determine the 
 leading terms in  the equations \rf{cur}, \rf{cury}. This could be done, in principle, 
 by reconstructing this effective action  from the superstring flat-space S-matrix for
  massive string states 
 using the NSR approach \ci{bl}.  This approach,however, 
    contains potential subtleties and we will 
 not follow it here.\foot{One of the subtle issues  (cf. \ci{bl})  
  is related to possible 
 mixing of string states with different masses in 3-point amplitudes and the
need  to understand all such mixings in order to extract the ``two massive -- two massless''
 4-point terms in the effective action.}
 
 Instead,  below we  will use  the ``semiclassical''   approach
 to computation of  energies  of ``short'' string states that was 
  initiated in \ci{tir}. It is based on the full \adss superstring action and thus 
     incorporates the fermionic contributions but it requires 
 certain assumptions of how to  interpret the semiclassical results, i.e. 
     how to  interpolate  them to  finite values of spins
    characterising proper  quantum string states.  
  
%expectat from vertex ops  ... but fermions not known..
%example of BL...  attempt   but there some problems  with it  doubts about correctness
%canonical dim are not seen at strong coupling ? well, seen in the shift...
%string level not seen at weak coupling ?   may be hidden in multiplet label... 
%ABA  --  looks like   wants to cancel canonical dim in $\g$ but this is only for *some* states. 
%Konishi: weak coupling series convergent, so what then to expect at strong coupling ? 
%Controlled by weak-coupling expansion.
%Strategy:
%$$ E = 2 \sqrt \sql [ 1 +   b/\sql ] + c $$
%$$c=\D_0 - 4 $$ 

%\renewcommand{\theequation}{2.\arabic{equation}}
% \setcounter{equation}{0}
%\section{General considerations}
% Expansions and their overlaps}
%\subsection{Structure of vertex operators}
%Comment on reproducing the expectation \rf{tri}....
%if there is $E$ term in subleading correction... 

\iffalse 
moral is that  if we get linear in E  term at next order 
 such 1/sql terms will appear...
but I though we should not really get such terms... 
a' corrections seem to be postponed... also that will be unpleasant  
for BPS  conditions -- we will need also J terms there to balance...
But...  this might be possible 
\fi

\renewcommand{\theequation}{3.\arabic{equation}}
 \setcounter{equation}{0}

\section{Energies of  quantum  strings   from semiclassical expansion}
% Expansions and their overlaps}

%\subsection{ Standard   semiclassical expansion.}

The standard  semiclassical  expansion  was extensively applied to the study 
of energies  of  strings in \adss  having  large quantum   numbers and thus 
  dual to ``long'' SYM operators 
with large  canonical dimensions (see, e.g., \ci{tser} for  reviews). 
It was suggested  in \ci{tir} that despite being formally valid   for ``large'' strings with large 
energies and spins  this expansion may be still useful  also  for extracting information about 
``small'' or  ``slow''   strings, 
assuming  that the resulting expressions for the energies admit analytic continuation to 
the region of small quantum  numbers such as spins. 
 In the cases we discuss below this  assumption  appears to be justified, i.e. 
 it is 
 consistent with 
 other  sources of information  about the structure of the 
 spectrum of quantum strings  in \adss.

Consider a classical string  solution  with energy $E$ and 
%some global charge (e.g., spin) 
spin  $J$. 
The standard semiclassical approximation   is based  on expanding $E$  in  large $\sql$  with 
$\J = { J \ov \sql} $ kept fixed, 
\be
 \la{sem}
 E= E( { J\ov \sql}, \sql)   =  \sql  \E_0 ( \J) +   \E_1 (\J)  + { 1\ov  \sql}   \E_2(\J) + ...  
\ee
In the ``short'' (or  ``slow'') 
 string limit  when $ \J \ll 1$    one  finds (cf. \rf{trh})
 %\foot{One may argue for this  structure 
% of quantum corrections   based on the expected analyticity  of 
 %the string partition function  in (mass)$^2$
% parameters of string fluctuations which are proportional to $\J + \OO(\J^2)$
% and the fact that to obtained $\E_n$ from 2d effective action one is to divide 
% by $\k \sim \sqrt \J$, see below.}
% \foot{Here in $\E_n$  we ignore
%where we included  
%a  constant $c_n= c_1 \delta_{n1}$  ($n \geq 1$) 
% which may  naively appear at least at  the 1-loop order but  should be absent  in a 
% proper treatment  of the semiclassical path integral
%  (see below). It may, however, appear in the 
%  exact expression  for the corresponding quantum string energy.}
%\foot{In general, $\E_1$ may 
%contain  also a $\J$-independent constant term which for simplicity 
%we shall ignore here (we will include it in the explicit discussions below).}  
\be \la{exxa}
&&\E_k=   
 \sqrt{  \J } \ ( a_{0k}   + a_{1k} \J + a_{2k} \J^2  + ... )  +  \E^{(\nan)}_{  k} \ , \\
 &&\E^{(\nan)}_{  k} =  c_{0k} +   c_{1k} \J  + ... \ .  \la{nea}
 \ .
\ee 
The ``analytic'' terms \ci{tir} written explicitly in \rf{exxa}
are the only ones present in the classical string energy and the 
ones that should naively appear from 
quantum corrections   if one assumes  analyticity  of 
 the string partition function  in (mass)$^2$
 parameters of string fluctuations (this follows from the fact 
 that (mass)$^2 \sim \J + \OO(\J^2)$ 
 and that to obtain $\E_n$ from the  2d effective action one is to divide 
it  by $\k \sim \sqrt \J$, see below).
The ``non-analytic'' terms in  $E_k^{(\nan)}$
%R
%$\E_{\nan \ k}$ 
originate from quantum ``infrared'' 
effects  in the small-spin  limit. 
%Here we included  constant term  $c_n= c_1 \delta_{n1}$  ($n \geq 1$) 
%which  may  naively appear at least at  the 1-loop order;
%it is actually ambiguous in the semiclassical treatment as we shall discuss below 
%but  should be absent  in a 
% proper treatment  of the semiclassical path integral
%  (see below). It may, however, 
%but should  indeed appear in the 
% exact expression  for the corresponding quantum string energy.
%\foot{In general, $\E_1$ may 
%contain  also a $\J$-independent constant term which for simplicity 
%we shall ignore here (we will include it in the explicit discussions below).} 

Formally, this  expansion   is   valid  for large $\sql$ and 
fixed $ \J= { J \ov \sql}$,  i.e. $ J \sim \sql  \gg 1$.
However,  if  we knew all the terms in it  to arbitrary order $k$
we could 
re-express  $\J$ in terms of $J= \sql \J $,   fix  $J$  to  certain {\it finite}
 value and then re-expand $E$    in  large $\sql$ for  fixed $J$. 
This is what one  would need to do in order  to compare with  gauge-theory
 results for short operators in the  strong coupling expansion.

Rewriting the above   expansion \rf{sem}
 in terms of $J$ we get 
\be \la{ex}
&&E= \sqrt{ \sql J}\ \Big[ a_{00}   + {  a_{10}  J + a_{01} \ov \sql } 
 + { a_{20} J^2  +  a_{11}   J + a_{02} \ov ( \sql )^2 } 
+ ...\Big] + E^{(\nan)} \ , \\ 
&&  E^{(\nan)}  =  {\cc}_{01} +     { {\cc}_{11}J  + {\cc}_{02}\ov \sql} + ...   \ , \la{nex}
\ee
where $a_{mk},\ c_{mk}$ are coefficients of the $k$-loop string sigma model corrections.
%and $\cc_1= c_1 + { c_2 \ov \sql} + ... $. 
If we now set $J$ to  some finite value  then  in order to know, e.g.,   the coefficient  of
the  $ 1\ov (\sql)^k$
term in the square bracket in   \rf{ex} % for such finite  $J$ 
  we  would  need to know only a {\it finite number}  of  coefficients  of up to  $k$-loop 
term in the semiclassical expansion \rf{exxa}.\foot{Let us  stress  that this is a remarkable feature
 of  the ``short string'' expansion, as compared to the 
``long'' or ``fast'' ($\J \gg 1$) string   expansion  considered in \ci{ft2}:
  there   the energy  expressed in terms of 
   $J$  contained the  tension  $\sql$  in positive powers 
so to  get a strong coupling expansion of the energy 
at  fixed $J$ one  would need to resum the whole  semiclassical series,
i.e. that  would  require one to 
know the infinite number of semiclassical coefficients.}
 
 For  example, 
  the knowledge of the 1-loop  coefficient $a_{01}$  
   together with the classical string energy  
coefficient  $a_{10}$ is sufficient to fix the 
$ 1\ov \sql$ term in the bracket in \rf{ex}.  To   fix the $ 1\ov (\sql)^2$
 term,    in addition to  
 the   classical and the 1-loop   corrections   one 
  would need to know also the 2-loop coefficient $a_{02}$, etc. 
  The same applies to the ``non-analytic''  part $E^{(\nan)}$.
  % given by  ${\cc}_{01} +     { {\cc}_{11}J \ov \sql} + ... $.

 Fixing a specific value of $J$ corresponding to some particular quantum string state 
  we then end up with the   strong-coupling expansion of the 
 energy (or dimension of the  corresponding  ``short'' 
 operator) already quoted in equations \rf{trh}, \rf{non}, \rf{aan}.
 % \be
% && E=   \sqrt{\sql } \Big[ k_0  + {  k_1 \ov \sql }  + { k_2  \ov ( \sql )^2 }  + \OO( { 1 \ov (\sql)^3}) 
%  \Big] +  {\cc}_1 \cr
% && \ \  \  = \   k_0  \fl  + c_1  + {k_1 \ov \fl } +    { k_2  \ov (\fl)^3 } +    \OO({ 1 \ov (\fl)^5 })   \ .  
% \la{qu} 
% \ee
 This is also the  same structure of the strong-coupling expansion of $E$  
 as  predicted 
 by the  consideration of  the marginality condition of the corresponding vertex operators, 
 see  \rf{fow}, \rf{ch}, \rf{maa}
 (in the notation of \rf{tri} $a_{00}\sqrt{J}  \to  2 \sqrt{n-1}, \ c_{01} = b_0$, 
 % \ c_1 = b_0, \ k_1 = b_0, \ k_2 = b_3, $  
 etc.). 
 
 In interpolating semiclassical expressions to finite values
 of spins we will need to take into account that, since
 we started in the region where  $J \gg 1$, we should ensure 
 that the resulting  expression for the energy  has the right flat-space limit as 
 appropriate for a quantum string state  with finite $J$; that may  require 
  to do a formal shift $J$   by a finite amount like $J \to J-2$.

Below we shall  consider  several explicit examples of expansions \rf{ex}
for simple string solutions that can be interpolated to quantum string 
states  that  carry the
 same quantum  numbers  as some of the bosonic members of the  
  Konishi multiplet from   Table  \ref{Ktable}.
% which we copy from \ci{ferr,bia}). 
We will include the classical and the  1-loop string corrections  and 
 verify  that,  as expected, the coefficient $b_1$  in \rf{non}, \rf{fow} is 
 %$k_0$ and $k_1$   in \rf{qu} 
   universal, while 
$b_0= c_{01}$   may change by integer shifts within the multiplet.

\subsection{Small circular spinning string with $J_1=J_2$  in $S^5$}

We shall start with  one of  the simplest  non-trivial 
 string  solutions in
\adss   --  a rigid circular  string rotating with two equal spins 
on   an (arbitrary-size)  3-sphere inside $S^5$. 
This is one of the two $J_1=J_2=J$ solutions found in \ci{ft2} -- the one  which is  stable and has 
$J  < \ha \sql$. The other (more well-known) one   has  $J  \geq  \ha \sql$ and describes 
a  string   rotating   on  a ``big'' (unit radius)  $S^3$ of  $ S^5$  
% (which has the same radius as $S^5$) 
   and is unstable
 against small perturbations.  
 
 The first (or ``small-string'')    solution  has classical energy  being of the same 
 form as in  flat space, 
 $E_0 = \sqrt{ 4 \sql \ J }$.  
 The second  (``large-string'')  solution   has  larger 
  energy $E_0 = \sqrt{ (2J)^2 + \l }$  for all   $J$
 apart from  the ``critical point''  $J = \ha   \sql $  where  the two  solutions coincide. 
 While  the second  string    is never small (it has radius of $S^5$)  and 
 admits a ``fast-string'' expansion  $\J = { J\ov \sql} \gg 1$, the first one 
  may  have an arbitrarily  small  radius and spin  and thus has 
 a  ``small-string'' limit  $\J \ll 1$  when 
    it probes the near-flat region   of $S^5$.

 In fact, the ``small-string'' solution is a direct 
 embedding into 
 %$R_t \times S^3$ part of 
 \adss    of the following flat-space  $R_t \times R^4$ solution describing a 
 rigid circular  string   rotating  in two orthogonal planes of   $ R^4$ \foot{Here $\sigma \in 
 [0, 2 \pi)$.  We shall always choose the  ``winding'' numbers to be 1.}
\bea \la{ff}
&&t=\kappa\tau \ , \ \ \ \ \ 
{\rx_x}\equiv  x_1+ i x_2 =\  a \ e^{i ( \tau +\sigma) }\ , \ \ \ \ \ 
 {\rx_y}\equiv x_3+ i x_4 =\  a\  e^{i  ( \tau -\sigma) }  \ , \\
&&
E_{\rm flat}  = {\te  {\kappa \ov \a' } } = \sqrt{ {\textstyle{4\ov  \a'}}  J }\ , \ \ \ \ \ \ \ \ 
 \ \  J_1=J_2=J =  {\te{  a^2 \ov \a'}} \ .  \la{yy} \eea
 Identifying the oscillator modes that are excited  on this solution one may associate  it  with the 
  quantum string state which is created by the following 
  %``target-space holomorphic'' 
   vertex operator
 (dots stand for the  fermionic terms generally present in the superstring case) 
 \be \la{vert} \ e^{-iEt}\ 
  \Big[(\del \rx_{x})^{J_1}  (\bd \rx_y)^{J_2} + ... \Big] 
  \ , \ \ \ \ \ \ \ \ \ \ 
   \a' E^2 = 2 ( J_1 + J_2 -2)  \ . \ee 
   In the  $J_1=J_2$  case the quantum-state analog of the classical 
   expression for the energy in  \rf{yy} is thus  found by a shift $J \to J-1$ 
 \be \la{fla} 
 E_{\rm flat}  =\sqrt{ {\textstyle{4\ov  \a'}}  (J-1) } \ . \ee
 Then   $J_1=J_2=2$  case   corresponds to a state on the first  massive string level $n=2$.

 Below we will be interested also in similar  semiclassical string states  in \adss 
 which   in the small-string limit  approach  the above flat-space solution  \rf{yy}. 
 This will allow  us to relate semiclassical results  to several 
   members of the Konishi  multiplet 
 should be dual to string  states at   the first excited  string level in
  the near-flat expansion of
 the \adss superstring \ci{gkp1,bia,tir}.

 There are three obvious choices for  how one may  embed the solution \rf{ff} into 
 \adss:  
 
 (i) the  two 2-planes may  belong  to $S^5$  leading to the 
 $J_1=J_2$ ``small-string'' solution;
 
 (ii) the two 2-planes  may belong  to $AdS_5$ leading to  a $S_1=S_2$ ``small-string'' solution;
 
 (iii)  one of the   2-planes  may  belong  to $AdS_5$
 and the other to $S^5$,  leading to  an  $S=J$ ``small-string'' solution.
 
 We will discuss these three cases in turn in this and the following two subsections. 
 Interpolated to finite   values of the spins $J=2,\ S=2$ the corresponding  string 
 states  will represent different members of the Konishi multiplet and
 this will allow us to verify the universality of the  strong-coupling expansion 
 of the  4-d anomalous dimension of the dual gauge theory operators. 
 
 \

 The  direct counterpart  of   \rf{ff} in  $R_t \times S^5$\  
 is described by  \ci{ft2} \foot{Here $X_k$  are the embedding coordinates of $S^5$, 
 $X^2_1 + ... + X^2_6 =1$\  (i.e. we use $X_k$ instead of  $n_k$ in \rf{scc}).  For
comparison, the ``large-string'' branch of the $J_1=J_2$   solution  \ci{ft2}
 is described  by 
$ X_1+iX_2= { 1 \ov \sqrt 2}   e^{i(w\tau+\sigma)}, \ 
X_3+iX_4={ 1 \ov \sqrt 2}   e^{i(w\tau-\sigma)}, \ 
X_5+iX_6=  0 $, where $w= 2\J  =\sqrt{ \kappa^2 -1} $ is arbitrary.
Notice that here we use different notation for $S^5$ embedding coordinates $X_k$ as 
compared to $n_k$ in \rf{scc}.}
\be
&&t=\kappa\tau~  ,\ \ \ \ 
  X_1+iX_2=a  \  e^{i(\tau+\sigma)}, \ \   \ \ X_3+iX_4=a  \  e^{i(\tau-\sigma)}, ~~~\ \ \
X_5+iX_6=  \sqrt{ 1- a^2 }, 
\cr
&&{\cal J}_1={\cal J}_2=a^2= {\k^2 \ov 4} = \J= { J \ov \sql}\ , \ \ \ \ \ \ \ \ \ 
E_0=\sql\  {\cal E}_0=\sql\ \kappa = \sqrt{4\sql J} \ . \la{eer} \eea 
Remarkably, the {\it exact}  expression for the classical energy  has  the same ``Regge'' 
form as in flat space \rf{yy} with $ {1 \ov \a'} \to \sql$ (we set the radius of $S^5$ to be  1).

The  quadratic  fluctuations of the \adss string action near this  homogeneous 
solution were 
discussed in \ci{ft2,ft3}. Here we use the  corresponding fluctuation frequencies 
to compute the 1-loop correction to the classical energy in \rf{eer}. 
In addition to 2 massless ``longitudinal''  bosonic modes 
one   finds 4 massive   fluctuations in $AdS_5$ directions with 
\be
\omega^2_n={ n^2+4{\cal J} }\ , \la{i}
\ee
 and 2 massless and 2 massive   fluctuations in $S^5$, with the latter having 
 \be
\omega^2_n{}_\pm  = n^2 + 4(1 - {\cal J}) 
                  \pm 2\sqrt{4(1 - {\cal J})n^2 + 4{\cal J}^2} \ .  \la{oi}
\ee
 The 4+4  fermionic modes  have the fluctuation frequencies 
% given by 
\be
\td \omega^2_n{}_\pm  = n^2 + 1 + {\cal J} 
                   \pm \sqrt{4(1 - {\cal J})n^2 + 4{\cal J}} \ . \la{loi}
\ee
The 1-loop   correction to the string energy  is given by 
$E_1=\frac{1}{\kappa} E_{\rm 2d}$, where $E_{\rm 2d}$ is determined by the logarithm of the 1-loop
partition function, $E_{\rm 2d} = - { 1 \ov {\cal T}} \ln Z_1$, \ ${\cal T}   \to \infty$. Thus 
\be
E_1&=&\frac{1}{\kappa} E_{\rm 2d} = \frac{1}{2\sqrt{\J}  } E_{\rm 2d} \ , \ \ \ \ \ 
E_{\rm 2d}= \frac{1}{2}\sum_{n=-\infty}^\infty \Omega_n
=\frac{1}{2}\Omega_0+\Omega_1+\Omega_2+\sum_{n=3}^\infty \Omega_n
 \ , \la{oo}\\
\Omega_n&\equiv &4\omega_n+ 2n +\omega_n{}_+ +\omega_n{}_-
         -4(\td \omega_n{}_++\td \omega_n{}_-)\la{om}  \ . 
\ee
Expanding in small $\J$    we find (we isolate $\Omega_{0},\Omega_{1},\Omega_{2}$ 
since  the expansion of generic $\Omega_{n}$ is singular  for $n=0, \pm 1, \pm 2$)
\be
&&\Omega_0=-4+8\sqrt{\cal J}-2{\cal J}-{\cal J}^2
%-{\cal J}^3
+\dots \ ,  \  \ \ \ \ \ \Omega_1=2-4\sqrt{\cal J}+5{\cal J}-\frac{437}{48}{\cal J}^2
            %+\frac{16727}{1152}{\cal J}^3
	    +\dots\ ,\la{ki}  \\ 
&&\Omega_2=-\frac{5}{3}{\cal J}+\frac{44}{27}{\cal J}^2
          % -\frac{44629}{15552}{\cal J}^3
	  	  +\dots\ ,\ \ \ \ \ \ 
\frac{1}{2}\Omega_0+\Omega_1+ \Omega_2=\frac{7}{3}{\cal J}-\frac{3445}{432}{\cal J}^2
         %+\frac{346819}{31104}{\cal J}^3 
	 +\dots \ ,\la{kii} \\
&&	 \sum^\infty_{n=3}\Omega_n =  q_1 {\cal J}+  q_2 {\cal J}^2+\dots \  , \la{tel} \\ &&
 q_1= -\sum^\infty_{n= 3}\frac{4}{n(n^2-1)}=-\frac{1}{3}
\  , \ \ \ \ \ \ 
q_2 = \sum^\infty_{n= 3}\frac{-28 + 87 n^2 - 79 n^4 + 8 n^6}{n^3(n^2-4)(n^2-1)^3}
=\frac{3121}{432}-6\zeta(3). \no
\eea
Here, as expected,   $(E_{\rm 2d})_{\J \to 0} \to 0$  since  the  solution shrinks to a point 
in the $\J\to 0$ limit. 
Note also that the $\sqrt \J$  contributions coming from $\Omega_0$ and $\Omega_1$
  cancel against each other,
implying the  absence of  the constant shift $c_{01}$ (cf. \rf{ex}, \rf{nex}) in the
 corresponding expression for $E_1$. 
 Also, the sum of  $\Omega_n$ does not contain $\J^{3/2}$
 term  so there is also no  ``non-analytic''  $c_{11} J\ov \sql $ term
in $E_1$  (cf. \rf{nea}, \rf{nex}).

  Explicitly, we find (cf. \rf{ex}) 
\be\la{en}
&&\E_1= \sqrt{\J}   +  a_{11} \J^{3/2}  +   \OO(\J^{5/2}) 
\ , \ \ \ \ \ \  \ \ \ \  a_{11}=   -\frac{3}{8}  - {3}\zeta(3)    \ , 
\la{onn}\\
&&E=E_0+E_1  =   \sqrt{4 \sqrt{\lambda} J}\ \Big[1+  \frac{1}{ 2\sqrt{\lambda}}
  + \frac{a_{11} J}{ 2\lambda}  + \OO( { J^2 \ov \l^{3/2} }) \Big]  \ , \ \ \ \ \ \ \ \
  E^{(\nan)}_1=0   \ .  \la{py}
\ee
This result is formally valid  in the limit when 
$\sql$ is first taken to be large   for fixed  $ \J = {J\ov \sql}$ and 
{\it then} $\J$ is  taken to be small  so that $ J \ll \sql$. However,  as discussed above, 
 we may  formally try to interpolate it to finite values of $J$. 
In that case  the $\frac{J}{\lambda}$ term  in \rf{py} is of the same order
 as a 2-loop correction 
which we will not compute  and so we  should ignore it here.

The same applies 
to the non-analytic term: there might in principle be  2-loop 
correction  producing $c_{02}$ term in \rf{nex} which is
of the same order as the (absent) 1-loop 
$ c_{11} J \ov \sql $ term;  we find it   very unlikely that  $c_{02} \not=0$.
Thus we conjecture that $b_2$ in \rf{aan} should  be zero. 

%It is instructive to 
Comparing  \rf{py} with the flat-space energy 
of  the quantum string state \rf{vert}  corresponding to the classical
solution \rf{ff}, \rf{yy},  i.e. with \rf{fla}, 
%\be
% $ \a' E^2 =  2 ( J_1  + J_2 -2)   = 4 (J-1)  
% $.
%  \ , \ \ \ \ \ \ \ \   J_1 = J_2 = J  \ .   \la{fl} \ee
%The  (bosonic part of the) corresponding   vertex operator is 
%$  (\del x_{12}  \bd x^*_{12})^{J_1/2} (\del x_{34}  \bd x^*_{34})^{J_2/2} e^{iEt}$, 
%where $x_{mk}= x_m + i x_k$. 
%so that  $J_1=J_2=2$ corresponds to a state on  the first massive string level. 
we conclude that in order to 
 interpret \rf{py} as a quantum string energy we should 
 %ensure the right flat-space limit, i.e. 
 shift $J$ as 
 %to match $J$ there should be shifted as
  in \rf{fla},  i.e. 
 $J \to J-1$. Then   
\be 
\la{gil}
E= 2\sqrt{ \sqrt{\lambda} (J-1) }\ \Big[1+  \frac{1}{ 2\sqrt{\lambda}}
  + \OO( { J \ov \l}) \Big]  \ . 
\ee
%Taking this into account and  setting 
Setting now  $J=2$   we end up with 
% get from \rf{py} the following generalization
 % of the leading-order flat-space expression:
\be 
\la{gel}
E= 2\sqrt[4]{\l}\ \Big[1+  \frac{1}{ 2\sqrt{\lambda}}   + \OO( { 1 \ov \l }) \Big]  \ . 
\ee
The reason  for this  choice of 
 $J=J_1=J_2=2$ is that such a state  belongs to the first 
 excited string level  and  the corresponding representation \rf{yo} \ 
 $(E,0,0; 2,2,0)$ or in Dynkin label notation \rf{dyn}\    $[2,0,2]_{(0,0)}$
is present  in the table \ref{Ktable}  of supersymmetry descendants of the singlet 
Konishi operator  $\Tr ( \bar \P_k \Phi_k)$. 
Indeed, there is one of such states  at  each of the 
  levels $\ell=4$  (with $\D_0= 2 + \ha \ell =4$), 
$\ell=8$  ($\D_0= 6$) and  $\ell=12$  ($\D_0= 8$),  i.e.
\be
 [2,0,2]_{(0,0)} \ : \ \ \ \ \   \D_0=4 \  (1) \ ;  \ \ \ \  \D_0=6 \  (1) \ ;\ \ \ \  \D_0=8 \  (1) \ .
 \la{gpp}
 \ee
The  $\D_0=4$   Konishi   state is  represented by  the operator $\Tr ([\Phi_1, \Phi_2]^2)$
 from the su(2) sector of the  SYM theory.
% while  the other two with higher  canonical dimension 
%should have  extra commutators  with $\bar \P_i$ and  $ \Phi_i$  which  increase dimension without
%increasing the  charges. 

%It is natural  to try to identify the quantum string state 
%in the representation $(E,0,0; J,J,0)$ with $J=2$ as
% corresponding to the supersymmetry descendant 
%of the singlet Konishi state.
%In the Dynkin label
% notation of \ci{bia} this is the state 
%$[2,0,2]_{(0,0)}$. Such   states are    present in the Konishi multiplet  table 
%  at the supersymmetry descendant 
%levels  with the canonical dimension 4   and the  dimension 8.
%The level 4 state is  represented by  the operator  from the su(2) sector of the 
% SYM theory, $\Tr ([\Phi_1, \Phi_2]^2)$,  which has    the same  $SO(6)$  charges.
%Then \rf{gel} should    give   prediction for   the strong-coupling expansion  for the  dimension
%of this   operator, with the leading-order term $2\sqrt[4]{\l}$ being the 
%well-known flat-space-based   expectation  \ci{gkp1}.

According to  \rf{gel}, the  universal coefficient $b_1$ in \rf{ch}   should then 
be equal to 1. 
%As for $\bb_0$, 
It is not clear a priori  which of the three  states in  \rf{gpp}
should be described by the  above semiclassical $J_1=J_2$ 
 string; the corresponding dimensions are
expected to be different only by the  constant $\D_0$  term in \rf{ch}. 
Since the  above circular solution appears to have  lowest energy for given spins 
we shall conjecture that it represents the  lowest-dimension state with $\D_0=4$. 
In this case  the value  of $b_0=0$ in \rf{gel}  (cf.\rf{aan}, \rf{fow})   translates into 
\be \la{bee} 
\bb_0 = -4  \ ,  \ee 
as already quoted in \rf{su}. Further evidence for   these values of $b_1$ 
and $\bb_0$ will be provided below.

\subsection{ Small circular spinning  string with $S_1=S_2$ in $AdS^5$}

As another closely related example  let  us now  
consider the counterpart of   the flat-space 
solution \rf{ff} when the 
circular spinning  string  rotates solely  in  $AdS_5$ \ci{ft2,art}.
In terms of the  $AdS_5$   embedding coordinates $Y_a$  we get (in the conformal gauge)\foot{Here
$Y_0^2 + Y_5^2 - Y_1^2 - Y_2^2 - Y_3^2 - Y_4^2 =1$;
  again,  we use 
different notation for the  embedding coordinates than in \rf{scc}: $Y_a$ instead of $N_a$.}
%(we again  choose the  winding numbers to be 1)
\be \la{spi}
Y_0 + iY_5 = \sqrt{1 + 2 r^2}  \  e^{i \k t} \ , \ \ \ 
Y_1 + iY_2 = r \  e^{i ( w \tau +  \s) } \ , \ \ \ 
Y_3 + iY_4 = r \  e^{i ( w \tau -  \s) } \   . 
\ee 
Here   $ r=\sinh \r_0 =  { 1 \ov 4}\k^2  , \ \  w^2 = \k^2 + 1 $  and    the energy and the spins are 
given by 
\be\la{sss}
E_0=\sqrt{\lambda}\ {\cal E}_0 , 
~~~S_1=S_2 =S= \sqrt{\lambda}\ {\cal S}   , \ \ \ \ 
{\cal S}= { 1 \ov 4} \kappa^2\sqrt{\kappa^2 + 1 }  , \ \ \ 
{\cal E}_0=\kappa + \frac{2 \kappa{\cal S}}{\sqrt{\kappa^2 + 1}} \  . 
\ee
This solution again admits a ``small-string''  limit ($\S\to 0$)\foot{This solution  is stable  for $\S \leq 1.17$
 \ci{ft2,ptt2}.}
in  which it  represents a small  circular string   rotating  in two orthogonal 
planes 
around its c.o.m.  in the central near-flat region of $AdS_5$. 
Its flat-space limit is thus again  given by  \rf{ff}.
%  as for the $J_1=J_2$ solution discussed above.

%The real  solution for $\kappa(\S) $  as the
In the  $\S= { S \ov \sql} \ll 1 $ expansion 
\be\la{kiu}
\kappa&=&2\sqrt {\cal S} -2{\cal S}^{3/2}+9{\cal S}^{5/2}+\dots \ , \ee 
%\cr
%\frac{1}{\kappa}&=&\frac{1}{2{\cal S}^{1/2}}
%+\frac{{\cal S}^{1/2}}{2}-\frac{7{\cal S}^{3/2}}{4}+\dots
%\ee
and  expressed in terms of $S=\sql \S$ the  classical energy becomes \ci{ft2}
 (cf. \rf{exxa}, \rf{ex}) 
%leads to the tree-level energy
\be\la{clas}
E_0=2\sqrt{\sqrt{\lambda}S}\ \Big[1+\frac{S}{\sqrt{\lambda}}
   - \frac{3S^2}{2\lambda}+ \OO({ S^3 \ov \l^{3/2}} )\Big] \ . 
\ee
Here in contrast to the $J_1=J_2$ solution \rf{eer} 
the classical  energy 
contains  non-trivial ``curvature''   corrections 
which 
%has more complicated expression  in terms of $S$ which is  
 modify  the leading-order flat-space ``Regge'' behavior.

The 1-loop correction  to  the energy  of this solution  was computed in \ci{ptt2}.  
Expanding  the fluctuation frequencies in small $\S$ 
it is straightforward to find the corresponding analogs of \rf{onn}, \rf{py}. 
In addition to 5+2 massless modes (2 of which are canceled by 
the  conformal-gauge ghosts) there  are 3 non-trivial  massive $AdS_5$ fluctuation modes 
with  the characteristic frequencies $\omega^{(i)}_n$ ($i=1,2,3$) 
 given by the solutions of the cubic equation \ci{ptt2}
%Bosonic frequencies (0505130): $5+2$ massless (the $+2$ are cancelled by ghosts)
%and 3 are solutions of:
\be
&&\omega_n^6 + c_1\omega_n^4+c_2\omega_n^2+c_3 = 0  \ , \ \ \ \ \ \ \ \ \ \ 
c_1= -8 - 10\kappa^2 - 3 n^2 \ , \\
&&
c_2=16  + 40  \kappa^2 + 24 \kappa^4 + 8 \kappa^2 n^2 + 3 n^4\ , \ \ \ \ \ 
c_3= - n^2 (n^2 - 4)(n^2 - 4 - 2 \kappa^2) \ . 
%C_2= -8 - 10\kappa^2 - 3 n^2.
\ee
The  4+4 fermionic frequencies are \ci{ptt2}
\be
\td \omega^2_{n}{}_\pm =
n^2 + 1 + \frac{5}{4} \kappa^2 
\pm \sqrt{4  n^2 +  \kappa^2    + 3 n^2 \kappa^2 + \kappa^4} \ . 
\ee
Then the analog of \rf{oo} is 
\be \la{inn}
E_1= { 1 \ov 2\k} \sum_{n=-\infty}^\infty \Omega_n \ , \ \ \ \ \ \ 
\Omega_n = 5n +\omega_n^{(1)}+\omega_n^{(2)}+\omega_n^{(3)}-4(\td \omega_{n}{}_+
+\td \omega_{ n}{}_-) \  .
\ee
The $\S \to 0$  expansion  gives 
(cf. \rf{kii}, \rf{tel})\foot{The 
${\cal S}\to 0$ expansion of $\sum^\infty_{n= 3}\Omega_n$ contains only integer
powers of ${\cal S}$.}
\be\la{jjk}
&&\frac{1}{2}\Omega_0+\Omega_1+\Omega_2=-4\sqrt {\cal S}
-\frac{7}{3}{\cal S}+4{\cal S}^{3/2}+\dots \ , \la{lll} \\
&&
\sum^\infty_{n= 3}\Omega_n=\sum^\infty_{n= 3}
\frac{4}{n (n^2-1)}\ \S +{\cal O}({\cal S}^{2})=\frac{1}{3}{\cal S}
+{\cal O}({\cal S}^{2}) \ . \la{kkk}
\ee
Again,  $E_{\rm 2d}= \ha \sum^\infty_{n= -\infty}\Omega_n$ vanishes in the 
$\S\to 0$ limit when  the string shrinks to a point. However, in contrast 
to the case of $J_1=J_2$ solution  here $E_{\rm 2d}$
approaches zero as a  square root of spin instead of linear function of spin, 
i.e.  naively there is a  ``non-analytic'' contribution coming from 
$\S^{1/2}$ (and $\S^{3/2}$)  term in \rf{lll}.
% instead of  $E_{\rm 2d}\sim  \J$.
%This produces a  constant shift in $E_1$. 
%While some large $n$ expansion of the coefficient of ${\cal S}^4$ may
%be obtained, its resummation does not seem clear/possible.
% The 
%${\cal S}\to $ expansion of $\sum^\infty_{n= 3}\Omega_n$ contains only integer
%powers of ${\cal S}$.
%
% all terms contains integer powers of ${\cal S}$; therefore all
% half-integer powers of ${\cal S}$ in ${\cal E}_1$ are controlled by 
% $\frac{1}{2}\Omega_0+\Omega_1+\Omega_2$. Unfortunately, this does not
% give much information, as the factor of \kappa^{-1} leads to mixing
% between integer and half-integer powers of ${\cal S}$.
%
Dividing  by $\k$ in \rf{inn}  and using \rf{kiu} appears to lead to\foot{This
 expression was independently found  by A. Tirziu.} 
\be\la{taa}
E_{1  } (?) 
%&=&\frac{1}{\kappa}(\frac{1}{2}\Omega_0+\Omega_1
%+\Omega_2+\sum_{n\ge 3}\Omega_n)\cr
= -2 -\sqrt{\cal S}+{\cal O}({\cal S}) \ . 
\ee
\iffalse
In contrast to \rf{onn} here we get a constant shift (i.e. $c_1=-2$ in \rf{ex}) 
which  originates from the $\sqrt \S$ term in \rf{lll}   coming from the fermionic 
$n=1$ mode contribution.
 Indeed, in contrast to  the corresponding expression \rf{kii} in the $J_1=J_2$ 
 case  here the sum of the lowest modes in \rf{lll} is {\it non-analytic} 
 in the spin $\S$  parameter. It is the non-analytic $\S^{1/2}$ and $\S^{3/2}$ terms in 
 \rf{lll} that lead (upon division by $2 \k$, cf.  \rf{kiu}) 
 to the  -2 and  $\OO(\S)$ term in \rf{taa}. 
 If we would  take \rf{taa} at face value, then 
putting together $E_0$ \rf{clas} and $E_1$ \rf{taa}  we would   get 
\be\la{ssa}
E=E_0+E_1=
%-2+2\sqrt{\sqrt{\lambda}S}\left(1+\frac{S}{\sqrt{\lambda}}+
%{\cal O}(S^2)\right)+\left(
%-\sqrt{\frac{S}{\sqrt{\lambda}}}+{\cal O}({\cal S}^{3/2})\right)
%\cr &=&
 2\sqrt{\sqrt{\lambda}S}\ 
\Big[1+\frac{2S-1}{2\sqrt{\lambda}}+{\cal O}({S\ov \lambda})\Big] - 2  \ .  
\ee
%While generated at 1-loop, the free 
%constant $(-2)$ does not really 
%carry the coupling constant dependence that would identify 
%whether requiring %\be%\lim_{S\rightarrow 0} E_1=0%\ee%is justified.
\fi
However, a more careful analysis  described  in Appendix  
implies that this $-2$ constant shift    
%the presence of non-analytic terms  in \rf{lll}
 is an artifact of the procedure 
of representing the  1-loop correction as a sum  of characteristic frequencies
and expanding  each frequency in small $\S$ separately.
% before doing the sum. 
Computing the 1-loop correction to 
2-d energy  as a combination of logarithms of determinants 
of the quadratic fluctuation operators  and then expanding the result in small $\S$ 
%does not produce such non-analytic terms. 
leads actually to the vanishing result for the 
 coefficient of the  leading non-analytic term $\sqrt \S$ in $E_{\rm 2d}$. 
Then instead of \rf{taa}     one finds  
% instead of  \rf{taa} and \rf{ssa} 
\be\la{assa}
&& E_1 = -\sqrt{\cal S}+{\cal O}(\S) \ , \la{ata}\\
&&E=E_0 + E_1 =  2\sqrt{\sqrt{\lambda}S}\ 
\Big[1+\frac{S- \ha }{\sqrt{\lambda}}+{\cal O}({S^2\ov \lambda})\Big]  +   \OO( { S \ov \sql}) 
 \ .  \la{sssa}
\ee
Notice that the  leading 1-loop  term in the $S_1=S_2$ case \rf{ata}
differs from the leading 1-loop term in  the $J_1=J_2$ \rf{onn} only by a sign and 
$\J \to \S$. One may try to attribute  this sign difference 
to the difference in 
the sign of the curvature of  $S^5$ and of $AdS_5$.

As  in the case of the  small $J_1=J_2$ string,  the flat-space counterpart  of this solution 
\rf{ff}   corresponds to  the  quantum  string state associated to 
 \rf{vert} with $J_i \to S_i$  and $S_1=S_2$. 
%described  by 
%\be 
%\la{fli}
% (\del x_{12}  \bd x^*_{12})^{S_1/2} (\del x_{34}  \bd x^*_{34})^{S_2/2} e^{iEt}\ , \ \ \ \ \ \ 
%\a' E^2 =  2 ( S_1  + S_2 -2)   = 4 (S-1)    \ .   \ee
Then  $S$ in \rf{sssa} should be redefined 
$S \to S-1$ to match  the flat-space limit \rf{fla} (cf. \rf{isa}) 
\be\la{isa}
E= 2\sqrt{\sqrt{\lambda}(S-1)}\ 
\Big[1+\frac{(S-1)-\ha}{\sqrt{\lambda}}+{\cal O}({S^2\ov \lambda})\Big]  +   \OO( { S \ov \sql}) 
%-2 
\ . \ee
%Eq. \rf{isa}  suggests that 
This suggests that in  the case of $S=2$, i.e.  the corresponding    string  state 
%the state $(E,2,2;0,0,0)$  
 belonging to the first excited 
level,   the  strong-coupling 
expansion of  its  energy  should thus  be 
\be\la{esa}
E=
  2\fl\ \Big[1+\frac{1}{2\sqrt{\lambda}}+{\cal O}({1\ov \lambda})\Big] 
   +   \OO( { 1 \ov \sql}) 
  %-2 
   \ . 
\ee
Here the subleading  ${\cal O}({1\ov \lambda})$ term in the bracket  and the last  ``non-analytic'' 
term  $\OO( { 1 \ov \sql}) $ term are sensitive to the 2-loop string corrections and thus 
beyond our reach. 
Remarkably, the two leading strong-coupling terms  in \rf{esa}
%up to the constant -2 shift, 
are exactly the same as in  \rf{gel} 
 found above  for the $J_1=J_2=2$   string state. 

This is perfectly   consistent with the expectation that the  $S_1=S_2=2$ state or 
$(E,2,2;0,0,0)$    should also belong to the  Konishi multiplet and thus should
have the same  anomalous dimension  as the state $(E,0,0;2,2,0)$ represented by 
the $J=2$ limit of the $J_1=J_2$ solution. 
Indeed, in the  Dynkin-label
%/$SU(2)$ 
 notation  \rf{dyn}  % \ci{bia}
this state  corresponds to $[0,0,0]_{(2,0)}$  and there are two of such states 
%such  representations   do  appear
   in  the Konishi multiplet table \ref{Ktable} 
 (cf. \rf{gpp})
\be
 [0,0,0]_{(2,0)} \ : \ \ \ \ \   \D_0=4 \  (1) \ ;\ \ \ \   \ \  \D_0=8 \  (1) \ .
 \la{ppg}
 \ee
The corresponding gauge theory operator with  
 $\Delta_0=4$   is
%$\bar \Phi_k  (D_{1+i2})^2  (D_{3+i4})^2 \Phi_k$. 
$ \Tr([ D_{1+i2}, D_{3+i4}])^2$ or $\Tr( F_{ 1+i2, 3+i4})^2$.~\foot{It belongs to a family
of field-strength operators 
 \ci{heis} conjectured in \ci{ptt2} to be related to $S_1=S_2$  semiclassical strings.}

It is natural to assume again that  the $S_1=S_2=2$ string state   
correspond to the   Konishi multiplet   member    with $\D_0=4$. 
Then the   resulting  values  of $\bb_0$ and $\b_1$  as  predicted by \rf{esa} are 
 the same as in  \rf{su}, \rf{bee}.\foot{As for the value  of $b_2$, 
 as already mentioned it receives  contribution both from the 1-loop  $ {c_{11}S \ov \sql} 
 $ term 
 and 2-loop term $c_{02}\ov \sql $ (cf. \rf{nex}), and their sum  may vanish
 due to underlying supersymmetry of the theory, as suggested 
 by the  remarks we made in the context of the  vertex operator approach in section 2.}

\subsection{ Small circular  spinning  string with $S=J$  in \adss }

Another embedding  of the 2-spin flat-space  solution \rf{ff}  into \adss  
is found by  considering one spinning plane being in $AdS_5$ and another -- in $S^5$.
%simple  homogeneous   string solution  one may consider is describing a rigid circular  
%string  rotating in  both $AdS_5$  and $S^5$. 
The  well-known  rigid  circular $(S,J)$ solution of this type 
\ci{art,ptt1}  where the string in $S^5$  is wrapped on a  big circle,  
 does not, however, 
 admit 
a ``small-string''  limit in which  the classical energy takes the   flat-space Regge form \rf{yy}.
However,   it is  easy to construct its close  relative 
 that does have the required   limit.

 To achieve this  one  is to  put  the circular string on a 2-sphere  of an arbitrary 
 radius  inside $S^5$.
   In terms of the $AdS_5$ and $S^5$   embedding coordinates we then
get (cf. \rf{eer}, \rf{spi})
 \be \la{mm}
&&Y_0 + iY_5 = \sqrt{1+r^2}  \  e^{i \k t} \ , \ \ \ 
Y_1 + iY_2 = r \  e^{i ( w \tau +  \s) } \ , \ \ \  \ \ w^2 = \k^2 +1 \ , \\ 
&& X_1 + i X_2  = a \  e^{i ( \tau -  \s) } \   , \ \ \ \ \ \ 
X_3+i X_4 = \sqrt{1 -  a^2}   \  . 
\ee
Here $r=\sinh \rho_0$ and $a=\sin\g_0 $ determine the size 
of the string in $AdS_5$ and $S^5$ respectively. 
The conformal gauge    conditions imply 
\be   (1 + r^2 ) \k^2 = r^2 ( w^2 +1) + 2 a^2  \ , \ \ \ \ \ \ \  \ \ \ \ \ 
     r^2 w = a^2     \  . 
\ee
%The spins are $ \S= r^2 w, \ \J= a^2 $
Thus for this solution one  has $\S=r^2w =\J=a^2 \leq 1 $, i.e. $S=J \leq \sql $. Also,    
$\E_0= (1+r^2)  \k =   \k +  { \S \k \ov \sqrt{ \k^2 +1 } } $, 
where $\k$ satisfies the equation 
$ \k^2 = { 2 \S  \ov \sqrt{ \k^2 +1} } +   2 \S$
which is readily solved.    

Explicitly, we find (cf.  \rf{kiu}, \rf{clas}) 
\be 
&&\k = %{\te {1 \ov \sqrt 2}} 
\sqrt{   \sqrt{ \fo + 2 \S} - \ha + 2 \S} = 2 \sqrt \S - \S^{3/2} + {15 \ov 4}
\S^{5/2} + ... \ , \la{tui}\\
 &&
 \E_0= 
 %{\te {1 \ov \sqrt 2}} 
 \sqrt{   \sqrt{ \fo + 2 \S} - \ha + 2 \S} \ 
 \Big( 1 + {  \S \ov  \sqrt{   \sqrt{ \fo + 2 \S} + \ha + 2 \S}} \Big) = 2 \sqrt \S +\S^{3/2}+ ... \ , \la{lpp}    \\
 &&
E_0=\sql \E_0 = 2\sqrt{\sqrt{\lambda}S}\ \Big[1+\frac{S}{2 \sqrt{\lambda}}
   - \frac{5S^2}{8\lambda}+ \OO({ S^3 \ov \l^{3/2}} )\Big] \ . 
 %2 \sqrt \S\ \Big[1  + {1 \ov 2} \S - {5 \ov 8} \S^{2} + ... \Big]  \ . 
 \la{trt}\ee
%%2 \S = \sqrt{ 1 + \k^2} ( \sqrt{ 1 + \k^2} - 1 )\ , \ \ \ \ \ \ 
%\sqrt{ 1 + \k^2} = \ha + \sqrt{ { 1 \ov 4} + 4 \S } 
 %$\k$ has explicit solution. {\bf needs a check   and expansion -- seems *simpler* 
%Note that this   solution  is thus algebraically  simpler 
%than the previous ``large-string'' $(S,J)$ solution of  \ci{art,ptt1}.
In the small-size  or  $\S=\J \to 0$ limit (when $w\to 1, \ r \to a\to 0$) this 
solution reduces to the flat-space  one \rf{ff}  with the energy taking 
  the   form   \rf{yy}. 
 
At  the $\S=\J=1$ point  (where $a=1, \ \k= \sqrt 3, \ w=2, \ r = \sqrt 2$) 
 this ``small-string'' $S=J$  solution coincides 
with the   ``large-string'' $S=J$ 
solution discussed in  \ci{art,ptt1}.\foot{For completeness,  
let us recall the form of the ``large-string'' solution of \ci{art}
(as above, we assume that the two  possible  winding numbers are equal to 1): \ 
$Y_0 + iY_5 = \sqrt{1+r^2}  \  e^{i \k t}  , $\ \   $ 
Y_1 + iY_2 = r \  e^{i ( w \tau +  \s) }  ,$  \  \ $  X_1 + i X_2  = e^{i ( \omega \tau -  \s) } , $
  where $w^2 = \k^2 +1 \geq 1$,\ \  $\S= r^2 w = \omega = \J$.
  Then   
  $\E_0= \k+  { \S \k \ov \sqrt{ \k^2 +1 } }$, where $\k(\S)$ satisfies 
  %$\sqrt{\k^2 +1} (\k^2  - \S^2-1) - 2\S=0
$ \k^2 = { 2 \S  \ov \sqrt{ \k^2 +1} } +    \S^2 +1$.
This  cubic equation for $\k^2$ admits two real  solutions for $\k$
(third one is unphysical): 
$
%\k^{(1)}= \sqrt{\S^2 -1},\ \  \ 
\k^{(1,2)} = \sqrt{1 + \ha \S^2 \pm  \ha \S \sqrt{8 + \S^2}} .$
% \ \ \ \ \ \ \ 
%\ \ \k^{(2)}  = \sqrt{1 + \ha \S^2 -  \ha \S \sqrt{8 + \S^2}}. $$ 
The first solution  is defined for
% $\S \geq 1$, the second -- for
 any $\S \geq -1$,  and the second  -- for  any  $\S \leq 1$. 
%In \ci{art,ptt1} (where  the large $\S$ expansion was discussed) 
%only the second branch was considered.
The   corresponding energies are 
$$
%&&\E^{(1)}_0= 2 \sqrt{\S^2-1} \ , \ \ \ 
  \E^{(1,2)}_0= \sqrt{ 1 + \ha \S^2  \pm \ha  \sqrt{ 8 +  \S^2} } \ 
 \Big[ 1 + {  \S \ov   \sqrt{ 2 + \ha \S^2  \pm \ha  \sqrt{ 8 +  \S^2} }  }   \Big].
 $$
 Only the first  branch which admit the large $\S$ expansion, 
 $\E^{(1)}_0 =  2 \S + {1 \ov \S} - { 5 \ov 4 \S^3} + ... ,
$
 was considered  in \ci{art,ptt1} (where the existence of this simple analytic expressions for the energy 
 %when winding numbers are equal to 1
  was not noticed). 
In  the small $\S$ expansion we get 
$  \E^{(1)}_0= 1 + \sqrt 2\  \S + { \S^2 \ov 4} - {\S^3 \ov 8 \sqrt 2} + ...   $ and 
% \ \ \ \   \ \ \ \  \ \ \ \ 
$\E^{(2)}_0= 1 - { \S^2 \ov 4} - {\S^3 \ov 4 \sqrt 2} + ... \ . 
$
This   solution thus  does not have the flat-space Regge asymptotics;  this is not surprising 
since here the string is wrapped   on  a big circle of $S^5$  and its tension    gives large 
contribution to the energy even for small spin. 
At the limiting point  $\S=\J=1$  the above  ``small-string'' solution \rf{mm} 
goes over to  the first branch of the ``large-string'' solution;
in particular, both energies become equal 
$\E_0= \E^{(1)}_0=  {3 \sqrt 3 \ov 2}$ (while $\E^{(2)}_0=0$ at $\S=1$). 
For $ 0 < \S < 1$ the energy of the ``small-string'' solution is always smaller than that of the 
``large-string'' one. 
}

%The corresponding flat-space solution is  again  the same as in  the previous two cases and
%corresponds to \rf{ff}, i.e. it is yet another (third) 
%embedding of the  same  flat-space solution \rf{ff}  
%of the circular 
%string  moving with two equal  spins on $S^3$ of arbitrary radius inside $R^4$, (i.e. here this 
%is formal  $w=1$ limit when $Y_1 + iY_2 = a \  e^{i (  \tau + i \s) } ,
% \ \  X_1 + i X_2  = a \  e^{i ( \tau - i \s) } $). 
%($w=1$, $S=J$) 
%It seems natural to expect  that this state belongs to $\D_0=6$ level. 

To compute the 1-loop correction to the energy of this solution 
 it turns out to be more efficient to use the path
integral approach in which (see  Appendix)
% for the calculation of 1-loop corrections to the
%target space energy. Thus, one should evaluate
\be
E_1=\frac{1}{2\kappa}\int_{-\infty}^\infty\frac{d\ww}{2\pi}
\sum_{n=-\infty}^\infty
\ln\frac{P_B(\ww,n,\S)}{P_F(\ww,n,\S)} \ . 
\label{E1pathint}
\ee
Here $P_B$ and $P_F$ are, respectively,  the bosonic and
fermionic characteristic polynomials, i.e. the equations 
$P_B(\ww,n,\S)=0$ and $P_F(\ww,n,\S)=0$ determine the characteristic frequencies. 
%equations, respectively.
%
%%%%AdS%%%%%%%%%
%\begin{eqnarray}
%0&=&(k_0-k_1)^2(k_0+k_1)
%\left[k_0^2-k_1^2+\frac{1}{2}(1-4{\cal
%S}-\sqrt{1+8{\cal S}})\right]^2\cr
%&&\times \left[(\ww-n)((\ww+n)^2-4)
%+(3-8{\cal S}-3\sqrt{1+8{\cal S}})\ww -(1-\sqrt{1+8{\cal S}})n\right]
%\end{eqnarray}
%%%%%S%%%%%%
%\be
% 0=(\ww-n)^3(\ww+n)^5((\ww-n)^2-4(1-\S))
%\ee
%%%%%%%%%%%%%%%%%
%
$P_B(\ww,n,\S)$ is found to be 
\begin{eqnarray}
&&\!\!\!\!
P_B =(\ww-n)^5(\ww+n)^6\Big[(\ww-n)^2-4(1-\S)\Big]
\Big[\ww^2-n^2+{\te {\frac{1}{2}}} (1-4{\cal
S}-\sqrt{1+8{\cal S}})\Big]^2\cr
&&~~~
\times \left[(\ww-n)[(\ww+n)^2-4]
+(3-8{\cal S}-3\sqrt{1+8{\cal S}})\ww -(1-\sqrt{1+8{\cal S}})n\right]~~. \la{jq}
\end{eqnarray}
The fermionic characteristic polynomial is more  complicated and we will
give  only the first few terms in its  expansion in $\S$:
%%%
%\begin{eqnarray}
%0&=&(\ww^2-(n+1)^2)^3(\ww^2-(n-1)^2)^3\cr
%&&\times\left[(\ww^2-(n+1)^2)(\ww^2-(n-1)^2)+(1-3\ww^2+4\wwn+n^2){\cal
%S}\right]+{\cal O}({\cal S}^2)
%\end{eqnarray}
%%%%
\begin{eqnarray}
P_F&=&[\ww^2-(n+1)^2]^3\ [\ww^2-(n-1)^2]^3\cr
&& \times\Big[[\ww^2-(n+1)^2][(\ww^2-(n-1)^2]
+8(1-3\ww^2+4\ww n+n^2){\cal S}\Big]+{\cal O}({\cal S}^2)\ . \la{qj}
\end{eqnarray}
At the next order in the small $\S$ expansion the three-fold
degeneracy is lifted to a two-fold one.  We have checked that at the
junction point $\S=\J=1$  the characteristic frequencies following
from the equations $P_B=0, \ P_F=0$   reproduce the ones of the ``large-string'' $(S,J)$ 
solution found in \ci{ptt1}.

While the characteristic polynomials are naturally functions of $\S$,
their roots, for low mode numbers ($n=-1,0,1$), turn out to depend
on $\sqrt{\S}$ for small $\S$. It is therefore important to analyze
these modes separately. 
A short calculation shows  that the $\S$ dependence of the contribution of
the  low-lying modes is
\be
\int_{-\infty}^\infty \frac{d\omega}{2\pi}\ln \frac{P_B(\ww,-1,\S)}{P_F(\ww,-1,\S)}
\frac{P_B(\ww,0,\S)}{P_F(\ww,0,\S)}
\frac{P_B(\ww,+1,\S)}{P_F(\ww,+1,\S)}={\cal O}(\S) \ , \la{lhs}
\ee
%is such that, the $\ww$-integral of its logarithm cannot yield an
i.e. it  does not  
%R
yield an $\sqrt{\S}$-dependent leading term.
% (see Appendix). 

Explicitly, the  leading $\S$ dependence of the corresponding part 
of the integrand in \rf{E1pathint}, i.e.  the integrand of 
%$\sum_{n=0, \pm 1}  \ln {P_B \ov P_F}$, 
 \rf{lhs},  is thus 
 %thus  linear in $\S$  and is 
found  to be linear in $\S$
\be
-\frac{72(\ww^2+1)}{(\ww^2-1)(\ww^2-4)(\ww^2-9)}\ \S~~. \la{esw}
\ee
The  denominators here may be associated to propagators of various
modes of the world-sheet theory. This defines the correct treatment of
the $\ww$-integral around these poles to be 
%R
given by the usual $i\epsilon$
prescription; equivalently, we may just ``Wick-rotate'' the integrand,
using the fact that it decays  sufficiently fast at large $\ww$.
As a result, the  $\ww$  integral of \rf{esw} 
turns  out to vanish identically. 

The leading small $\S$ dependence of a generic term in the sum in
 (\ref{E1pathint}) may also be extracted by expanding the
integrand. 
For a generic term with $|n|\ge 2$ we get 
\be
- \frac{8\left[3 \ww^6 + 5 \ww^4 (3 n^2-1)
 - \ww^2 (15 n^4 - 76 n^2 + 32)-(n^2-4)^2 (3 n^2-1) \right]}
{[\ww^2 - (n-2)^2][\ww^2 - (n-1)^2] (\ww^2 - n^2) 
 [\ww^2 - (n+1)^2][\ww^2 - (n+2)^2]}\ ~\S~~.
\label{generic}
\ee
The apparent small $\ww$ singularity at $n=\pm 2$ is,  in fact,  cured by
the numerator, which is proportional to $\ww^2$ at those points. 
The absence of singularities in the integration domain of $\ww$ 
justifies this term-by-term expansion and confirms the absence of lower-order terms which are 
non-analytic in $\S$.

%With the same observation as in the case of the $n=-1,0,1$ modes
Defining the integral through  ``Wick rotation''  as discussed 
 in the case of the $n=-1,0,1$ 
modes,
implies that the integral 
of (\ref{generic}) also vanishes identically.
%R
%
This
vanishing may be confirmed by the direct analysis of the sum of the  characteristic 
frequencies for $|n|\ge 2$  (cf. \rf{tel}, \rf{kkk}).

All this  implies that the leading term in the small $\S$ expansion of the 
 integral in (\ref{E1pathint}) is proportional to $ \S^{3/2}$;  after
  dividing by $\k = 2 \sqrt \S+...$ in \rf{tui} 
    we conclude that here (cf. \rf{onn}, \rf{ata}) 
\be
E_1={\cal O}(\S)={\cal O}(\frac{S}{\sql})   \  . \la{qqq}
\ee
One  may try  to attribute the cancellation of the
 leading $\sim \sqrt \S$ 1-loop correction 
to the cancellation  between  the $AdS_5$ and $S^5$ contributions
(recall  the opposite signs of the 1-loop $\sqrt \J$ term in the $J_1=J_2$ \rf{onn}
and the 1-loop $\sqrt \S$ term in the  $S_1=S_2$  \rf{ata}  cases).

We conclude  that  the two leading terms
% $\sim \sqrt{ S} (1 + S + ...)$ 
 in the 
 1-loop corrected  energy of the small rigid circular $S=J$ 
string rotating both in $AdS_5$ and $S^5$  are   given  simply by the 
classical expression \rf{trt}.
For the corresponding quantum string state in the near-flat limit  we should 
find,  as in \rf{vert}, 
%The    flat space string level equation is then 
$\a' E^2 = 2 ( S+ J -2) =  4(S-1)$. 
Shifting $S \to S-1$ to  have the correct flat-space  limit  we end up with
(cf. \rf{gil}, \rf{isa}) 
\be 
E=  2 \sqrt{\sql(S-1)} \ \Big[1  + {(S-1)\ov 2\sql } +   \OO({ S^2 \ov \l})  \Big] 
+  \OO( { S \ov \sql})   \ . 
 \la{trta}
\ee
Since we are interested in a state  in the Konishi  multiplet that should belong 
to the first excited string level we should   interpolate this result to   $S=J=2$.
Remarkably, the  leading coefficients in  \rf{trta} for $S=2$  are  again the same as in the 
 $J_1=J_2=2$ \rf{gil}, \rf{gel}    and $S_1=S_2=2$  \rf{isa}, \rf{esa} cases
 discussed above: 
\be\la{iff}
E= 2\fl\ \Big[1+\frac{1}{2\sqrt{\lambda}}+{\cal O}({1\ov \lambda})\Big]+  \OO( { 1 \ov \sql})   \ . 
 \ee
This is again consistent  with the expectation that all these  states should 
belong to the same  supermultiplet, so their energies may  differ only  by a constant
$\l$-independent shifts.

The  representation corresponding to the $S=J=2$ state is  $(E,2,0; 2,0,0)$
or in the Dynkin label notation  $[0,2,0]_{(1,1)}$. 
There are 1+3+1  such  states   present
in the Konishi multiplet  table  \ref{Ktable}
%at the following  descendant levels 
(cf. \rf{gpp}, \rf{ppg}) 
\be
 [0,2,0]_{(1,1)} \ : \ \ \ \ \   \D_0=4 \  (1) \ ;\ \ \  \ \  \  \D_0=6 \  (3) \ ; \ \ \ \ \  \D_0=8 \  (1) \ .
 \la{ggg}
 \ee
The  dual gauge theory operator at level $\D_0=4$  is the familiar  one from the sl(2) sector:
$\Tr [\Phi_1 (D_{1+i2})^2  \Phi_1]$.  

We shall assume again that the $S=J=2$  state  is dual to 
the lowest-dimension  $\D_0=4$ state in the Konishi multiplet. Then in addition to
$b_1=1$ as  implied  by \rf{iff} we again find $\bb_0=-4$ 
as in the two previous cases.

\iffalse
Again, the  coefficient of the first subleading term   in \rf{iff}
will be doubled as in \rf{chm} 
if we process \rf{iff}  via $E^2 \to E(E-4)$  substitution. 
\fi

%%%%%%%%%%%%%%%%%%%%%%%%%%%%%%%%%%%%%%%%%%%%%%%%%%%%%%%%%%%%%%%%%%%%%%%%%%%%%%%%%%%%%

\subsection{Small folded  spinning  strings  in $AdS_5\times S^5$
%: \ $S$, \ $J$, \ and  $S=J$
 }

%\subsubsection{  $S$  }

One   may also consider  other  semiclassical  solutions
in \adss that  in the small spin  limit reduce to 
flat-space solutions that may be  interpreted  as   
representing  massive string states.
% at the  first excited string level. 
One familiar example is the rigid folded  string   in $AdS_5$ with spin $S$ \ci{dev,gkp2}.
There is a similar folded string solution in $S^5$   with spin $J$ \ci{gkp2}.
One  may also consider their  generalization when folded string is rotating both in 
$AdS_5$ and $S^5$  with spins  $S=J$. 
Interpolated  to  $S=4$, $J=4$ and $S=J=2$ respectively 
these  configurations   should represent different 
states at the first excited string level  and thus 
should be dual to different  states in the Konishi  multiplet. 

As we shall discuss   below, the 1-loop corrected energy for the corresponding \adss 
solutions when interpolated to the respective values of the spins 
reproduces the same  expression \rf{gel}, \rf{esa}, \rf{iff} 
as found above in the case of the circular string examples. 
This provides  further evidence of  the consistency of the suggested picture.

\subsubsection{\small Folded string with spin $S$  in $AdS_5$}

The  small spin limit of the classical   energy of the folded 
spinning string  in $AdS_5$  has the expected  behavior $E_0 = \sqrt{ 2 \sql S} + ...\ $.
The small-spin  
expansion of the 1-loop correction to
its   energy  is  more complicated to compute 
 than in the 
homogeneous string examples considered above as here the solution involves
 elliptic functions.
This problem was  first addressed 
 in \ci{tir}  and then also discussed  in an unpublished work in
\ci{gr,btt}.\foot{A generalization 
to include dependence on the string 
center-of-mass  momentum $J$ in $S^5$ was considered  in \ci{bet}.}
 The general structure of the quantum-corrected  energy found in the semiclassical expansion 
 ($\sql \gg 1, \ \S= {S \ov \sql}$=fixed) 
 and then expanded  in   $\S \to 0$   is   the same  as in \rf{ex}  (with $J$ replaced by $S$) 
 \ci{tir,btt} 
\be\la{ssai}
&& E= \sqrt{2\sqrt{\lambda} S}\ 
\Big[1+\frac{{\te{ 3 \ov 8}}   S  + a_{01}}{\sqrt{\lambda}}+\frac{ {\te - {21\ov 128}} S^2 
 + a_{11} S + a_{02} }{{\lambda}}   +  
{\cal O}({S^3\ov (\sql)^3})\Big] 
+  E^{(\nan)} \ , \\ 
&&  E^{(\nan)}  =  {\cc}_{01} +     { {\cc}_{11}S  + {\cc}_{02}\ov \sql} + ...   \ , \la{fex}
\ee
where the coefficients  $a_{01},a_{11},{\cc}_{01}$ are the 1-loop ones,  
$a_{02},{\cc}_{11}$ are the 2-loop one, etc. 
%(not computed in \ci{tir}), etc. 
The 1-loop values found in \ci{tir}  were 
$   a_{01} = 3-4 \ln 2 =0.227, \ \    a_{11} = - {1219\ov 576} +
 {3\ov 2} \ln 2 + {3\ov 4}\zeta(3)$.\foot{It
is interesting to note  that the presence of $\zeta(3)$ in the $a_{11}$ coefficient 
appears to be a universal feature -- it is also present 
 in the case of the $J_1=J_2$   string in 
\rf{onn}, \rf{py}. It should thus appear in the next-to-next-to 
leading coefficient $b_3$ in the strong-coupling
expansion  \rf{tri}, \rf{ch}
of the   anomalous dimension of the Konishi  operator.} 
An  alternative  computation of the leading 1-loop coefficient  $a_{01}$ in \ci{gr}
  (based on extracting 
the fluctuation spectrum from the algebraic curve description \ci{curve})   
led to a different  numerical 
result $a_{01}\approx - 0.25$. 
Due to some uncertainties  in the  treatment of the zero modes in the
 original computation in \ci{tir}, here we shall assume
that the result  of \ci{gr} is actually  the right  one, and in fact,
 is exactly given by\foot{At this order of
perturbation theory  the reasoning based on what one should expect to find 
by  computing the  anomalous  dimensions of the corresponding vertex
operators suggests that this coefficient  should be expressed 
in terms of rational numbers only.}
\be\la{aaa}
a_{01}=- { 1 \ov 4}  \ . \ee
Also, 
the analysis \ci{btt} of the separate zero-mode contributions
 (coming from  the mixed bosonic modes in $AdS_5$) 
  appears to   give 
  \be\la{bba}  \cc_{01}=2   \ . \ee
%%%%%
 % The analysis of the separate zero-mode contributions
 % (coming from  the mixed bosonic modes in $AdS_5$) 
 %  in \ci{btt} appears to lead, like in  \rf{lll},   to an additional   
 %  contribution $\delta E_1 = 2 +  \OO(\S)$. However, 
 %  we believe   that the separate small  spin expansion of the lowest 
 % mode frequencies  leading to non-analytic  in $\S$ terms is not the correct procedure
 % (see the discussion below \rf{ssa}, section 3.3 and Appendix).}
%%%%%% 
Assuming the validity of 
 \rf{aaa} and  \rf{bba}   the classical plus the  1-loop result for the energy is then found to be 
 \be\la{dsa}
E= E_0+E_1= \sqrt{2\sqrt{\lambda} S}\ \Big[1+\frac{{\te{ 3 \ov 8}}  
 S -\fo }{\sqrt{\lambda}}+{\cal O}({S^2\ov \lambda})\Big]   +  2  +  \OO( { S \ov \sql}) 
\ . \ee
The flat-space limit of this solution 
%is  the folded string rotating in a plane 
(cf. \rf{ff})
\bea \la{ftf}
t=\kappa\tau \ , \ \ \ \ \ 
{\rx_1}\equiv  x_1+ i x_2 =\  a  \sin \s \   e^{i  \tau  }\ , \ \ \ \ \ 
% {\rx_2}\equiv x_3+ i x_4 =\  a\  e^{i  ( \tau -\sigma) }  \ , \\
E_{\rm flat} = \sqrt{ {\te{2\ov  \a'}}  S }\ , \ \ \ \ \ \ \ \ 
 \ \  S=  {\te{  a^2 \ov 2\a'}} \ ,   \eea
is a semiclassical  counterpart of the quantum  string state  on  the leading Regge trajectory 
represented by the  vertex operator (cf. \rf{vert})\foot{The corresponding (bosonic) 
Fock space state is 
  $( a^\dagger_1  \td a_{1} ^\dagger)^{S\ov 2}  |0,E>$. The semiclassical 
  string is represented in this Fock space
  as a coherent state 
 exp$( \sqrt S a^\dagger_1
+ \sqrt S \td a_{1} ^\dagger) |0,E>$.} 
\be e^{-iEt}\ \Big[(\del \rx_x  \bar  \del \rx_x )^{S\ov 2} + ...\Big] \ ,  \ \ \ \ \ \ \ \ \ \ 
\a' E^2 = 2 (S-2)  \ . \ee
To be able to continue   \rf{dsa} to  small values of $S$ 
and match the correct flat-space limit one should    shift $S\to S-2$,
 thus  getting (cf. \rf{gil}, \rf{isa}, \rf{trta}) 
 \be\la{dsan}
 %v2
E= \sqrt{2\sqrt{\lambda} (S-2) }\ \Big[1+\frac{{\te{ 3 \ov 8}} 
 (S-2) -  \fo}{\sqrt{\lambda}}+   
{\cal O}({S^2\ov \lambda})\Big]    +     2  +  \OO( { S \ov \sql})       
\ . \ee
Then for the state on the first excited string level, i.e. for $S=4$, we 
finish with 
\be\la{usa}
E= 2\fl \Big[1+\frac{1}{2\sqrt{\lambda}}+{\cal O}({1\ov \lambda})\Big] 
+ 2  +  \OO( { 1 \ov \sql}) 
\ . \ee
 Remarkably, the first two leading terms here  are  exactly the same 
 %(modulo the constant shifts) 
 as in all the  three of the above circular string 
 cases, \rf{gel}, \rf{esa}, \rf{iff}.

This is  how  it should be 
%not surprising 
as the $S=4$ state should also belong to the Konishi multiplet and thus
 should have the same anomalous dimension. 
 The corresponding representation is 
 $(E,4,0;0,0,0)$ or  $[0,0,0]_{(2,2)}$ 
and there is indeed  just one  such state  in the Konishi multiplet table  \ref{Ktable}
(cf.  \rf{gpp}, \rf{ppg}, \rf{ggg})\foot{The  dual 
SYM operator should  contain terms like   $\bar \Phi_k  (D_{1+i2})^4 \Phi_k$.}
\be
 [0,0,0]_{(2,2)} \ : \ \ \ \ \   \ \ \   \  \D_0=6 \  (1)  \ .
 \la{igg}
 \ee
 Since this state has $\D_0=6$,  the constant  shift $b_0=2$  in \rf{usa} (cf. \rf{aan}) 
 is then perfectly consistent with the value of $ \bb_0=-4$ in
\rf{su}, \rf{bee}.

% If we again apply the same $E^2 \to E(E-4)$ recipe    as in the $J_1=J_2=2$ case we will again
% end up with the quantum state  energy \rf{chm} and strong-coupling coefficients 
% \rf{esu}.

%The +2 difference in constant shifts between the $J_1=J_2=2$ state \rf{gel} and the $S=4$ state 
% \rf{usa}  can be then attributed  to the difference in the  corresponding 
% descendant levels (i.e.  canonical dimensions at weak coupling):  $\D_0=4$ vs 
% $\D_0=6$.
% [However, the $-2$ shift in the case of the  $S_1=S_2=2$ state \rf{isa} which 
%  should also   have 
 % $\D_0=4$ (or $\D_0=8$) is a puzzle....
%  The shift in the  $J=S=2$ case should shed some light... ]
 
\subsubsection{\small Folded string with spin $J$ in $S^5$ }

Similarly to case of the flat-space  circular string   \rf{ff}
  that can be embedded  either in $S^5$ 
or in $AdS_5$ (or both) we can also embed the flat-space folded 
string \rf{ftf}   not in $AdS_5$ but in  $S^5$. 
The corresponding solution \ci{gkp2} is 
the direct analog of the one in $AdS_5$.\foot{Explicitly, 
in terms of
the embedding coordinates of $S^2$ inside of  $S^5$ we have  
 $X_1+ i X_2 = \sin \psi(s) \ e^{i w \tau}, \ \ X_3 = \cos \psi(s), \ \ \psi'^2  +  w^2  \sin^2 \psi=\k^2$.}
In that case the classical energy has the following  
small $\J= { J \ov \sql}$ expansion 
\be\la{bk}
E_0= \sqrt{2\sqrt{\lambda} J}\ 
\Big[1+\frac{{\te{ 1 \ov 8}} J}{\sqrt{\lambda}}+{\cal O}({J^2\ov \lambda})\Big] \ .  \ee
While the 1-loop correction  in this case  was not computed  so far, 
we shall   conjecture that the  coefficient $a_{01}$ in the analog of \rf{ssai} 
here should  have the {\it opposite} sign  compared to \rf{aaa}  since the sign of the 
curvature of $S^5$ is opposite 
to that  of $AdS_5$. Indeed, as we have seen   
on the examples of the $J_1=J_2$ and $S_1=S_2$   circular string solutions, 
the respective 1-loop coefficients in \rf{gil} and \rf{isa}  differ  only  by  the sign. 
%We may then
We shall thus assume  that for the folded string in $S^5$ 
one  should get $a_{01}= {1 \ov 4}$.
% instead of     \rf{aaa}.
We shall also  assume that the constant 
$ \cc_{01} $  in the corresponding  analog of the ``non-analytic'' part
of the 1-loop energy \rf{fex} should be again given by 2 as in  \rf{bba}.
 %\be\la{bba}  \cc_{01}=2   \ . \ee

Taking also into account the shift $J \to J-2$
to match the required  flat-space  limit   we 
can then   generalize \rf{bk} to the following 1-loop corrected result  (cf. \rf{dsa})
\be\la{bki}
E=E_0+E_1= \sqrt{2\sqrt{\lambda} (J-2)}\ 
\Big[1+\frac{{\te{ 1 \ov 8}} (J-2)  + \fo }{\sqrt{\lambda}}+{\cal O}({J^2\ov \lambda})\Big]
 + 2 +   {\cal O}({J\ov \lambda})
 \ .  \ee
%By analogy with \rf{dsa} we also assumed the same constant shift $+2$. 
We observe   that  for  the  state with 
$J=4$ which is at  the first excited level the  value  of  \rf{bki} 
is the same as in   \rf{usa}, i.e.   gives  $b_1=1$ as  in all other cases   discussed above.
%i.e.  gives  $b_1=1$. 
  
The state with $J=4$  is in  the representation 
$(E,0,0;4,0,0)$  or $[0,4,0]_{(0,0)}$;  there  is    just one such state 
in the Konishi table \ref{Ktable} (cf. \rf{gpp}, \rf{igg}):\foot{The   
 SYM operator dual to it  may  contain terms like 
 %of have the structure 
  $\Tr [\P_1,[\P_1,\bar \Phi_k]   [\P_1, [\P_1, \Phi_k]]$.}
 \be
 [0,4,0]_{(0,0)} \ : \ \ \ \ \   \ \ \   \  \D_0=6 \  (1)  \ .
 \la{img}
 \ee
 % The corresponding SYM operator may have the structure 
  % $\Tr [\P_1,[\P_1,\bar \Phi_k]   [\P_1, [\P_1, \Phi_k]]$.
 %To reproduce  the same   dimension  as 
  %  the $S=4$ state providing further  evidence of consistency 
 % of the emerging picture. 
 As in the  previous folded string example, the $b_0=\D_0 + \bb_0=2$ is
 then again consistent   with  $\bb_0=-4$.

%%%%%%%%%%%%%%%%%%%%%%%%%%%%%%%%%%%%%%%%%%%%%%%%%%%%%%%%%%%%%%%%%%%%%%%%%%%%%%%

\subsubsection{ \small  Folded string with two spins  $S=J$ in \adss }

%{ \bf Folded spin with two spins  $S=J$  }

Finally, as in  the third ``mixed'' embedding of the circular string in \adss 
discussed in the  subsection 3.3, 
we may consider also another   $(S,J)$   solution   given by 
the direct  superposition of the  folded strings  rotating in 
$AdS_5$ and in  $S^5$ ``glued'' together by the Virasoro condition. 
Here the leading terms in the small-spin expansion of 
the classical energy  take the expected   ``direct  superposition'' 
form  (cf. \rf{ssai} and \rf{bk})\foot{This
 follows from the 
straightforward combination of the 
folded string solutions  in $AdS_3$ and in $R \times S^2$ \ci{gkp2}. 
 We thank A. Tirziu for the  derivation of this expression.}
\be \la{sj}
E_0 = \sqrt{2\sqrt{\lambda} (S +J)}
\ \Big[1+\frac{{\te{ 3 \ov 8}}  S +{\te{ 1 \ov 8}} J }{\sqrt{\lambda}}+{\cal O}({S^2\ov \lambda})\Big] \ . \ee
%%Then \rf{isa} would suggest that 
In particular, for $S=J$ the leading two terms here 
become  exactly the same  as in  the energy of the small circular $S=J$ 
string \rf{trt}  discussed above:
%AT
\foot{Let us mention that there is yet another  familiar  $(S,J)$ string    obtained 
giving the  folded string in $AdS_5$ an angular momentum $J$ in $S^5$  \ci{ft1}.
In this case the small-spin limit of the classical  energy is \ci{ft1,bet}
$$E_0 = \sqrt{2\sqrt{\lambda} S  +  J^2 }\ 
\Big[1+\frac{{\te{ 3 \ov 8}}  S }{\sqrt{\lambda}}+...
%{\cal O}({S^2\ov \lambda})
\Big]
= \sqrt{2\sqrt{\lambda} S } \ \Big[1+\frac{{\te{ 3 \ov 8}}  S  
+  {J^2 \ov 4S} }{\sqrt{\lambda}}+{\cal O}({S^2\ov \lambda})\Big] $$ 
%Then \rf{isa} would suggest that 
%This agrees with naive expectation
%$E(E-4) =  2\sqrt{\lambda} S  +  J(J+4) + ... $. 
One may expect that the  corresponding    state  on the first excited string level 
should  than still have $S=4$ as in the $J=0$ case. 
 The corresponding representation  $(E,4,0; J,0,0)$  or 
$[0,J,0]_{(2,2)}$   is not, however,  in the Konishi multiplet  table for $J > 0$ 
so we will not discuss this case here.}
%In fact, the only $[a,b,c]_{(2,2)}$  state there   is 
%$[0,0,0]_{(2,2)}$  at $\D_0=6$. 
%But then   how we identify the Konishi in sl(2) sector? 
%Usual option $\Phi D^2 \Phi$ seems $S=2, J=2$. 
%But this is
%}
\be \la{sji}
E_0 = 2\sqrt{\sqrt{\lambda} S}
\ \Big[1+\frac{S }{2\sqrt{\lambda}}+{\cal O}({S^2\ov \lambda})\Big] \ . \ee
The flat-space limits of the  two $S=J$ solutions are, however, 
 different -- the 
circular $S=J$ string \rf{mm} reduces to   \rf{eer}
 while 
the folded $S=J$  string  still reduced to the folded string
rotating in {\it one}  plane  \rf{ftf}.\foot{Indeed, the $S=J$ folded string in \adss in  the flat limit 
is described by $x_1+ i x_2 = a \sin \s  e^{ i \tau}, \ \ x_3 + i x_4 = a \sin \s  e^{ i \tau}$, 
so by rotation  $x'_1 = {x_1 + x_3 \ov \sqrt 2}, \ \ x'_2= {x_2 + x_4 \ov \sqrt 2}$
this is still equivalent to a  folded  string spinning only in one plane $(x'_1,x'_2)$
with  spin $S'=2S$.}

In the circular $S=J$ case the leading $\sqrt S$  term in the 
1-loop correction $E_1$ happened to cancel  out  (see \rf{qqq})
and we interpreted this as a  cancellation of the 1-loop corrections  in \rf{onn} and in \rf{ata}
if we could   formally put them  together.
If we assume  that the { leading}
1-loop correction  in the folded $S$-string \rf{dsa} and the folded $J$-string \rf{bk}  energies 
can also  be directly  superposed  (as it is the case for
%like  this is so  for 
%R
the classical contributions in  \rf{sj})
then the total  1-loop coefficient in the analog of \rf{ssai}  would 
 be  $a_{01} = -{ 1 \ov 4} + { 1 \ov 4}=0$, 
i.e.   it would   vanish just like  in the circular $S=J$ case. 
Then we  would   finish  with the following result for the 1-loop corrected energy 
(after shifting  $S+J =2S \to 2S -2$  
%or $S\to S-1,\  J \to J-1$
 to make \rf{sji} match the flat-space limit) 
\be \la{sjii}
E = 2\sqrt{\sqrt{\lambda}( S-1)}
\ \Big[1+\frac{(S -1)}{2\sqrt{\lambda}}+{\cal O}({S^2\ov \lambda})\Big]
 + 2 + \OO( { S \ov \sql})    \ . \ee
Here we assumed the same ``non-analytic'' constant term as in the other two 
 folded string cases \rf{usa} and \rf{bki}. 
 
Modulo the constant $+2$  shift this  happens to be exactly the same expression  \rf{trta}
as found earlier  in the circular $S=J$ case. 
Then the  choice of  $S=J=2$ gives again a state on the   first excited string level. 
%as in the circular $S=J$ case  
and  \rf{sjii}  reproduces   the same expression for the first two leading terms 
 in \rf{iff}  as in all other previous cases. 
%anomalous dimension as in the other cases. 

The  representation corresponding to the folded $S=J=2$   state is 
the same as in the circular $S=J=2$ case, i.e. 
$(E,2,0; 2,0,0)$ or $[0,2,0]_{(1,1)}$. 
 There are  5  such states 
in the   Konishi table \ref{Ktable}  already  listed in \rf{ggg};  we repeat them again 
here\foot{As 
was already 
mentioned below \rf{ggg}, the  operator  dual to the $\D_0=4$ state should  be  
the familiar sl(2) sector  one    $\Tr [\Phi_1 (D_{1+i2})^2  \Phi_1]$. 
The operator  for the $\D_0=6$  state may  contain terms like 
 $\Tr [\bar \Phi_k, D_{1+i2}\Phi_1]  [\Phi_k,   D_{1+i2}\Phi_1]$, etc.}
%(cf.  \rf{gpp},\rf{ppg},\rf{ggg},\rf{igg},\rf{img}):
\be
 [0,2,0]_{(1,1)} \ : \ \ \ \ \   \D_0=4 \  (1) \ ;\ \ \  \ \ 
  \  \D_0=6 \  (3) \ ; \ \
  \ \ \  \D_0=8 \  (1) \ .
 \la{gggi}
 \ee
%One may conjecture  that  both  the circular string  $S=J=2$ state and this folded $S=J=2$
%string   state should both  belong to $\D_0=6$ level. 
%???  No, folded string should have less energy and thus belongs to 
%$\D_0=4$ level ? 
Given that we identified the circular $S=J=2$ state with a $\D_0=4$ state in 
 \rf{gggi}, it is natural to assume that  the folded $S=J=2$ state, 
 like   the folded $S=4$ \rf{igg}  and $J=4$ \rf{img} states, 
 should correspond to one of the  three  $\D_0=6$  states in representation 
  $[0,2,0]_{(1,1)}$   in the 
 Konishi multiplet table. 
 
 The proposal is then that  the  three circular  solutions 
  represent Konishi states at level $\D_0=4$
 while the three folded  solutions  represent Konishi states at level $\D_0=6$.
 This appears  to be  in line with each of these   two groups of solutions
 having distinct flat-space limit (cf. \rf{ff} and \rf{ftf}).

\iffalse
\subsubsection{Various}
Folded string in $AdS_5 $ with extra  c.o.m.  $J$ in $S^5$:
\be 
&&E_0 = \sqrt{2\sqrt{\lambda} S  +  J^2 }\ 
\Big[1+\frac{{\te{ 3 \ov 8}}  S }{\sqrt{\lambda}}+...
%{\cal O}({S^2\ov \lambda})
\Big]
\cr
 &&
= \sqrt{2\sqrt{\lambda} S } \ \Big[1+\frac{{\te{ 3 \ov 8}}  S  
+  {J^2 \ov 4S} }{\sqrt{\lambda}}+{\cal O}({S^2\ov \lambda})\Big] \ . \ee
%Then \rf{isa} would suggest that 
This agrees with naive expectation like:
$E(E-4) =  2\sqrt{\lambda} S  +  J(J+4) + ... $. 
The corresponding  Konishi state should be $S=4$  (to get right level) 
and $J=2$.  The corresponding representation is $(E,4,0; 2,0,0)$  or 
$[0,2,0]_{(2,2)}$  but it is not in the Konishi table ?! 
In fact, the only $[a,b,c]_{(2,2)}$  state there   is 
$[0,0,0]_{(2,2)}$  at $\D_0=6$. 
But then   how we identify the Konishi in sl(2) sector? 
Usual option $\Phi D^2 \Phi$ seems $S=2, J=2$. 
But this is not on the right string  level, unless R-charge is represented by 
extended string not   by c.o.m as naively  assumed. 
Pulsating string:
\be 
E(E-4) = L(L+4)  + \l { L^2 - J^2 \ov 2 L^2}   + ... \ee
$L= J + N$, \ \ 
$E\E= \sqrt{ 2 N} \ \Big[  1 - \fo N + ... \Big] $
\fi

%%%%%%%%%%%%%%%%%%%%%%%%%%%%%%%%%%%%%%%%%%%%%%%%%%%%%%%%%%%%%%
\section{Summary}
%%%%%%%%%%%%%%%%%%%%%%%%%%%%%%%%%%%%%%%%%%%%%%%%%%%%%%%%%%%%%%%%

As we have argued above,  the interpolation of semiclassical  expressions for 
1-loop corrected energies   of  two classes of  spinning string solutions to 
small values of spins  corresponding to quantum string states at the first excited level 
leads to the following expression (cf. \rf{ch}, \rf{su}, \rf{cog}) 
\be \la{um}
E= 2 \fl + \D_0 - 4 +   {1 \ov \fl} +  \OO( { 1 \ov (\fl)^3})  \ .  \ee
Here  $\D_0=4$ for the  three  states in the Konishi multiplet  table 
 $[2,0,2]_{(0,0)}$ \rf{gpp}, $[0,0,0]_{(2,0)}$  \rf{ppg},
 and  $[0,2,0]_{(1,1)}$ \rf{ggg}
 represented by  the three circular string configurations,    and 
$\D_0=6$ for the  three  states  
$[0,0,0]_{(2,2)}$  \rf{igg}, $[0,4,0]_{(0,0)}$  \rf{img} and   $[0,2,0]_{(1,1)}$ \rf{gggi}
represented by the three folded  string configurations.
The universality of the coefficients in $E-\D_0$ is consistent with
the expectation that all gauge-theory states in the same
supermultiplet should have the same anomalous dimension.
%
\iffalse
It also lends strong support to the validity of our proposal. Indeed,
while the various solutions discussed in this paper should be related
by symmetry transformations, these symmetries are not manifestly
realized in the GS action and not maintained throughout the
calculations; this makes the symmetry algebra sensitive to the
regularization scheme. The universality of the coefficients in
$E-\D_0$ found here is a nontrivial confirmation that our methods
indeed realize the $psu(2,2|4)$ symmetry algebra at 1-loop level. 
\fi
It also lends strong support to the validity of our proposal. Indeed,
%none of 
the $psu(2,2|4)$ generators that could relate the various
solutions discussed in this paper are not manifestly realized in the 
quantum theory  based on the  GS action. Their realization at 
the quantum level is highly
dependent on the choice of a regularization scheme. The universality of
the coefficients in $E-\D_0$ found here is a nontrivial confirmation
that our methods indeed realize the $psu(2,2|4)$ symmetry algebra at
1-loop level.
%vvv2

In   \rf{um} we conjectured  that 2-loop coefficient $b_2$ in \rf{aan} vanishes 
so that the leading correction  to the first three leading terms in strong-coupling expansion 
is  determined  by the ``analytic'' 2-loop term of order $ { 1 \ov (\fl)^3}$. 

It is interesting to note that \rf{um}  has very  similar form 
to the expansion of  energy of a massive scalar in $AdS_5$ 
with a mass corresponding to the first excited string level (cf. \rf{eex})
%AAT
\be  E(E-4) = m^2_0= 4 \sql   \la{eqa} \ ,  \ee
i.e. 
\be 
E=  2  +  \sqrt{4  \sql  +  4  }   
=  2 \fl +  2  +   {1 \ov \fl} +  \OO( { 1 \ov (\fl)^3})  \ . 
\la{cos} \ee
Heuristically,  one may argue that  the mass of the corresponding 10-d scalar  should not receive 
leading $ \a' = {1 \ov \sql}$ correction since a candidate for 
the leading background-dependent correction in the case of the scalar 
operator  \rf{cur} vanishes for \adss background.
% (the curvature scalar and ${\rm F}_5 {\rm F}_5$ for \adss background vanish) 
The constant 2 in   \rf{cos} would  be consistent with \rf{um}
if the corresponding scalar  would be dual to the $\D_0=6$ state  in the Konishi multiplet. 
There are indeed three singlet  $[0,0,0]_{(0,0)}$ states with $\D_0=6$  in the Konishi multiplet table 
\rf{Ktable},  but the significance of this observation remains to be understood.

%AAT
As follows from our discussion in section 2.2,   interpreting  $E$ as  the solution of 
the marginality  condition for the corresponding string vertex operators, the first two subleading  coefficients $b_0$ and $b_1$ in \rf{tri}  must be rational 
because  the 1-loop 2-d anomalous dimensions may  contain only rational  coefficients.
At the same time, the semiclassical string computations in \ci{tir} and here \rf{onn}
imply that  $b_3$ should already be transcendental, containing $\zeta(3)$. These 
are robust predictions of our approach. 

At the same time, 
one may wonder if there might be some subtlety in our interpretation 
of a semiclassical result for the string energy 
$E$ interpolated to small values of spins as directly 
representing the quantum string energy.\foot{For example, one may wonder if one 
may  need to  shift $E$ 
by an integer just as we did shift  spins to  match the flat-space limit.}
One  may suspect   that our semiclassical result for $E_0+E_1+...$ 
(let us denote it $E_{\rm sc}$) computes, in fact, the quantum-corrected 
 string mass $m=m_0 +...= \sqrt{ \sql (n-1)}+ ...=E_{\rm sc}$. Then to get the value 
 of the quantum string $AdS_5$  energy  one would 
  need still to solve the equation like \rf{eqa}, 
 i.e. $E_{\rm q} (E_{\rm q}-4) = E_{\rm sc}^2$. It is easy to see that  in this  case 
 the value of the coefficient $b_1$ for a state on the first  excited level 
 will double from 1  in $ E_{\rm sc}$ in  \rf{um}   to 2  in $E_{\rm q}$.
 To match the right values of $b_0$ for different states in the Konishi multiplet
 one will  need 
 to use a more complicated  ansatz like $E_{\rm q} (E_{\rm q}-4p_1) +p_2 = E_{\rm sc}^2$
 (with $p_1,p_2$=const).
 This  prescription, however, seems  ad hoc, 
 so we hesitate to  advocate it here.

 Still, intriguingly, $b_1=2$  appears to be the value coming out of the very recent 
 numerical solution for the strong-coupling expansion 
 of the dimension of the Konishi operator  from  integrability (Y-system) approach \ci{gkv}.

%$E$ is not $\Delta$ ? 

\iffalse 

\section{Comments}

Massless    higher spins in $\l\to 0$ limit? seems problematic to take
$\l\to 0$ limit from string (strong-coupling) side since
inverse tension expansion does not have finite radius of convergence. 
> A thing that crossed my mind is that perhaps it may help to look at a (JJJ)
> solution
> -- if any is available, since there are [00,2]_{0,0} reps at D0=3. But in
> the ent we will run in the same problem as with the 2-J solution... This
> representation appears at higher levels too... so we won't be able to say
> which solution sits where...
that  is possible but will get progressively more messy -- 3-spin solution
-- unless we take limit  J_1=J2, J_3 or rep  [J2-J3,0,J2+J3]
is hard. I guess it would be nice to see  which shifts we get in this case.
May  be this is not that bad at the end as we had frequencies in
papers with Frolov and with F and Park.

open problems -- way to deal with dets;  3-spin cases; 
pulsating string as dual to a scalar.

\fi

\section*{Acknowledgments }
%%%%%%%%%%%%%%%%%%%%%%%%%%%

We thank  M. Beccaria,  S. Frolov,  N. Gromov,    R. Metsaev, A. Rej, F. Spill 
and  
% especially    N. Gromov and  
 A. Tirziu  for  many  useful  discussions.
We would like to  thank N. Gromov    for sharing with us some unpublished results. 
 AAT  also thanks  A. Tirziu for collaboration on some  related problems. 
RR's work was supported by the US Department of Energy under contract
DE-FG02-201390ER40577 (OJI), 
the US National Science Foundation under grant PHY-0608114 and 
the A. P. Sloan Foundation.
 Part of this work was done while  we were  participants of the 2009 
program ``Fundamental Aspects of Superstring Theory'' 
at the  Kavli Institute for Theoretical Physics at Santa Barbara.  
Our work there was supported  in part by the
 National Science Foundation under Grant No. PHY05-51164.
%(28  March - 18  April 2009).
AAT  also acknowledges the hospitality of the Galileo Galilei Institute in Florence 
during the  2009 program  ``Non-Perturbative Methods in Strongly Coupled Gauge Theories".
%(25 April- 8 May, 2009).

%\appendix{  }  

%%%%%%%%%%%%%%%%%%%%%%%%%%%%%%%%%%%%%%%%%%%%%%%%%%%%%%%%%%%%%%%%%%%%%%%%%%%

%\def \cC {{\cal C}}

\appendix
%\addcontentsline{toc}{section}{Appendices}
%\addcontentsline{toc}{section}{Appendices}
\section*{Appendix: Path integral approach to 
computation \\
of 1-loop correction to string  energy 
}
\refstepcounter{section}
\def\theequation{A.\arabic{equation}}
\setcounter{equation}{0}

%\subsection{Path integral approach to 1-loop energy shift }

As discussed at length in, e.g.,  \cite{ft2,ftt, rt0712}, loop corrections to 
energy of classical solutions may be efficiently evaluated in the 
%conformal gauge using 
  path integral approach  in the conformal gauge.
The
1-loop correction to the energy of a classical solution (soliton) of
the world-sheet theory is given in terms of the logarithm of the 
determinants of the kinetic operators of the bosonic and fermionic 
quadratic fluctuations around the solution:
\be
E_1=\frac{1}{2\kappa }\int_{-\infty}^\infty\frac{d\ww}{2\pi}
%\sum_{n=-\infty}^\infty
\ln\frac{\det K_B }{\det K_F }~~.
\label{E1pathint_det}
\ee
We are assuming that the solution is stationary in $\tau$
(with $t = \kappa \tau$) so that  the determinants 
are 1-dimensional ones. 
In  the closed string case   where  the theory defined on a  spatial cylinder 
%A presentation of this contribution, which makes accounts
they can be expressed  
%for the finite extent of the worldsheet space-like direction, is 
in  terms of the characteristic polynomials, $P_B(\ww,n,\cC)$ and
$P_F(\ww,n,\cC)$, of the bosonic and fermionic fluctuations
%respectively:
\be
E_1=\frac{1}{2\kappa }\int_{-\infty}^\infty\frac{d\ww}{2\pi}
\sum_{n=-\infty}^\infty
\ln\frac{P_B(\ww,n,\cC)}{P_F(\ww,n,\cC)}~~.
\label{E1pathint_gen}
\ee
Here $\cC$ denotes some 
%(unspecified) 
charges (rescaled  by string tension, $\cC = {{\rm C} \ov \sql}$) 
%(s densities)
characterizing the classical solution;  $\kappa$ is also a  function of them 
through  
the conformal gauge conditions. 
%defines the
%relation between the worldsheet and target space time coordinates
%\be
%t=\kappa(\cC)\tau~~.
%\ee
%Its dependence on charges is determined by the e\cCuations of motion
%and the Virasoro constraints. 
%It is moreover the case that,
For each value of $n$, the $\ww$ integral is convergent at large
$\ww$  (the string sigma-model is UV finite). 

The roots of $P_B$ and $P_F$  are the usual
characteristic frequencies. If the characteristic polynomials
$P_B(\ww,n,\cC)$ and $P_F(\ww,n, \cC)$ factorize into  products
\be
\prod_I [\ww-\omega_I(n,\cC)][\ww+\omega_I(n,\cC)]
\ee
then the $\ww$ integral may be trivially carried out and one obtains
the standard expression for $E_1$ as a sum over characteristic
frequencies $ \omega_I$.

The dependence on the charges $\cC$ should  be extracted from the
expressions (\ref{E1pathint_det}) and (\ref{E1pathint_gen}) with 
 care. Since the charges  $\cC$ are  parameters of the
classical solution, they appear analytically in the quadratic
fluctuation Lagrangian and thus  in the characteristic equations.
 The roots
of the characteristic equation may,  however,
% and thus the $\ww$ integral,
depend on fractional powers of $\cC$, e.g., on  $\sqrt{\cC}$. This
may  occur for a finite set of mode numbers $n$.
We may thus distinguish the two types of contributions: the  analytic in $\cC$ and
 the non-analytic in $\cC$.
 % They may be treated separately.

To find  the analytic contributions 
 one may consider
evaluating the determinants in (\ref{E1pathint_det}) or the $\ww$
integral in (\ref{E1pathint_gen}) in  a perturbative expansion in
$\cC$. This amounts to 
interpreting as perturbations 
all the   terms in the string 
quadratic fluctuation Lagrangian  which depend
 on the parameters of the classical
solution. 
%(in particular,  all off-diagonal terms of the bosonic 
%propagator).
%\footnote{
This expansion  is thus analogous to the mass insertion formalism
in 4d QFT.\footnote{The GS fermions should be treated with care 
%in this formalism, 
since their entire kinetic term  may be  proportional to some
charge. In this case one is 
% One may, e.g., 
to redefine the fermions to absorb the leading charge dependence.}

%This approach also exposes
 The  presence of  non-analytic
$\cC$-dependence 
%-- if any is present. 
%Indeed, such dependence
%marks a breakdown of the assumption of analyticity (which allows
%one to carry out an expansion in the charge densities) and 
will  manifest 
itself as a breakdown of this perturbative treatment. 
In particular, it may happen that 
at some order in small  $\cC$ expansion, the $\ww$ integral will be divergent at finite
values of $\ww$.\foot{The integral over $\ww$ is  to be convergent
at $\ww\rightarrow\pm\infty$ due to UV finiteness.}
 By carrying out the expansion of the 
terms in the summand in equation (\ref{E1pathint_gen}) one may be able 
to identify the mode numbers responsible for potential 
 non-analytic
terms.\footnote{It is possible that additional non-analytic terms may 
arise from a resummation of the modes.} 
The corresponding fractional power of the
charge will lie between the integer powers of $\cC$ for which the last 
convergent and the first divergent $\ww$ integrals may  occur.

The values of $n$ for which the singularities in the small $\cC$ expansion
may occur should  be analysed separately. While a priory  the fractional
powers of $\cC$ could  appear at high orders in the small $\cC$ expansion, in
all the  cases we discussed above they potentially  occur as the
leading term, even before the first analytic term. 
%In such cases, and
Assuming  the characteristic equations have the symmetry
$(\ww,n)\leftrightarrow (-\ww,-n)$, the leading $\sqrt{\cC}$ dependence
can then be easily extracted by a simple change of variables in  the $\ww$
integral. Namely, we are to consider all the apparently  singular terms
(labelled by $n_s$) 
together 
\be
\ln\prod_{n_s} \frac{P_F(\ww,n_s,\cC)}{P_B(\ww,n_s,\cC)}~~.
\ee
%where  $n_s$  label  finitely many such modes. 
The assumed symmetry of the characteristic equations  guarantees that this
logarithm is a real function and that the $\ww$ integral is well
defined.
% Analyzing this in detail allows, through a change of
%variables, the extraction of the leading nonanalytic dependence on
%charges.

For the circular (homogeneous)  
  rotating string solutions discussed in sections 3.1-3.3
%{\bf section number} 
the potential   non-analytic dependence on the spins $\S$ or $\J$ 
arises from   factors of the type
\be
\ln \frac{[(\ww-n_0)^2-\cC]^m}{(\ww-n_0)^{2m}} \ , 
\ee
where $m$ is  some even integer.\footnote{The power $m$  
 is even due to the assumed symmetry 
$(\ww,n)\leftrightarrow (-\ww,-n)$.} 
 Changing the variable $\ww$, we can then 
extract the $\sqrt{\cC}$ dependence of the 1-loop correction to the energy. 
The coefficient of the $\sqrt{\cC}$  term  is given by  a well-defined
integral which  happens to vanish identically.
% (see the discussion in section 3.3).
 %vanishes exactly. 
% It is important to realize
%that the even value of $m$ is crucial for having a well-defined
%integral and a real value for the 1-loop 
%correction to the energy. 
%Indeed, if
%$m$ were odd, then the logarithm would develop an imaginary part upon
%integration over all values of $\ww$. 

%AT
As an example, let us consider in some detail the case of the small 
circular  string in $AdS_5$ with $S_1=S_2$  discussed   in section  3.2.
 As one can check by 
analyzing the characteristic equations, the only potential non-analytic 
contributions arise from the modes with  numbers $n=\pm 2, \pm1, 0$.\footnote{One may 
see this by simply expanding the argument of the logarithm at small $\S$ 
and noticing  the appearance of singularities for finite values of $\omega$.}
Combining these modes together as
\be
\ln\prod_{n=-2}^2\frac{P_F(\ww,n,\S)}{P_B(\ww,n,\S)}
\label{pr}
\ee
and using the explicit form of the characteristic polynomials, we find 
that  to extract the leading $\S\to 0$ dependence of the integral of  \rf{pr}, 
the argument  of the logarithm  in \rf{pr}
 can be  simplified to\footnote{This essentially amounts to dropping all 
$\S$-dependence that does not introduce singularities in the $\omega$ integral. 
Such terms necessarily yield  only subleading  ${\cal O}(\S)$ contributions.}
\be
\prod_{n=-2}^2\frac{P_F(\ww,n,\S)}{P_B(\ww,n,\S)} \ \ \to \ \ 
%\!\!\!\!\!\!\!\!
 && \frac{[(\omega-1)^2-\S]^4}{[(\omega-1)^2]^4}
\frac{ (\omega^2-\S)^8}{(\omega^2)^8}
\frac{[(\omega+1)^2-\S]^4}{[(\omega+1)^2]^2} \cr 
&&\ \ \ \ \ \  \times \frac{[(\omega-1)^2]^2}{[(\omega-1)^2-2\S]^2}
\frac{[(\omega+1)^2]^2}{[(\omega+1)^2-2\S]^2}
 \ . \la{simp}
\ee
Indeed, it is clear that naively expanding (\ref{pr}) and (\ref{simp})
 at small $\S$ leads to singular $\omega$ 
integrals. 

Splitting the logarithm of (\ref{simp}) into the sum of 
logarithms of the factors shown above,   allows, through simple 
changes of variables in the $\omega$-integral of each of the resulting 
terms, to extract the leading $\sqrt \S$ dependence as
\be
\sqrt{\S} \ \int_{-\infty}^\infty \frac{d\omega}{2\pi}
\ln\frac{ (\omega^2-1)^8}{(\omega^2)^8}~~.
\ee
The  integral here vanishes (as can be seen explicitly by carefully writing the 
integral as a combination of  simple logarithms and shifting the integration variable), 
 implying the vanishing of the coefficient  of the leading non-analytic 
 $\sqrt \S$ term in the 1-loop correction to the energy. 

%AT
In general, it  would be important to clarify the  structure of the small spin expansion 
and the issue of analytic and non-analytic terms in  the 1-loop 
corrections   similar to the one discussed above further, e.g., 
using other methods of evaluating the 1-loop determinants
or    attempting to do  the summation over modes  before expanding 
in the  small-spin parameter. 

\iffalse
 In particular, one may consider 
It is nice that you bring up the point with analytic/non-analytic terms in
appendix. Just for completeness it might be that there are also exponential terms
like e^(-1/\S), if one does the small \S expansion unambiguously (i.e. doing full
sum/integral and then expanding in small \S). For the circular J_1=J_2 solution
this can probably be checked by doing the exact sum and then expanding.
It could be that there are also ln S terms in the non-analytic part in the case
of folded string -- we have seen such possible terms with Beccaria when
including J -- but maybe with no J such terms are not there, I am not sure.
\fi

%%%%%%%%%%%%%%%%%%%%%%%%%%%%%%%%%%%%%%%%%%%%%%%%%%%%%%%%%%%%%%%%%%%%%%%%%%%%%%%%%%%%%%%%%%%%%%%%%%
\begin{table}[htb] 
\centering
{\footnotesize
\begin{tabular}{|c|l|} \hline
     $\D_0$ & $[p_1, q, p_2]_{(s_L, s_R)}=
[J_2-J_3, J_1-J_2, J_2+J_3]_{( {S_1+S_2 \ov 2}, {S_1-S_2 \ov 2})}$\\ \hline\hline
     $2$& $
     [0,0,0]_{(0,0)}
     $ \\ \hline $
     \dDelta+\frac{1}{2}$& $
     [0,0,1]_{(0,\frac{1}{2})}
     +[1,0,0]_{(\frac{1}{2},0)}
     $ \\ \hline $
     \dDelta+1$& $
     [0,0,0]_{(\frac{1}{2},\frac{1}{2})}
     +[0,0,2]_{(0,0)}
     +[0,1,0]_{(0,1)+(1,0)}
     +[1,0,1]_{(\frac{1}{2},\frac{1}{2})}
     +[2,0,0]_{(0,0)}
     $ \\ \hline $
     \dDelta+\frac{3}{2}$& $
     [0,0,1]_{(\frac{1}{2},0)+(\frac{1}{2},1)+(\frac{3}{2},0)}
     +[0,1,1]_{(0,\frac{1}{2})+(1,\frac{1}{2})}
     +[1,0,0]_{(0,\frac{1}{2})+(0,\frac{3}{2})+(1,\frac{1}{2})}
     +[1,0,2]_{(\frac{1}{2},0)}
     $\\& $
     +[1,1,0]_{(\frac{1}{2},0)+(\frac{1}{2},1)}
     +[2,0,1]_{(0,\frac{1}{2})}
     $ \\ \hline $
     \dDelta+2$& $
     [0,0,0]_{(0,0)+(0,2)+(1,1)+(2,0)}
     +[0,0,2]_{(\frac{1}{2},\frac{1}{2})+(\frac{3}{2},\frac{1}{2})}
     +[0,1,0]_{2(\frac{1}{2},\frac{1}{2})+(\frac{1}{2},\frac{3}{2})+(\frac{3}{2},\frac{1}{2})}
     +[2,0,2]_{(0,0)}
     +[2,1,0]_{(0,1)}
     $\\& $
     +[0,1,2]_{(1,0)}
     +[0,2,0]_{2(0,0)+(1,1)}
     +[1,0,1]_{(0,0)+2(0,1)+2(1,0)+(1,1)}
     +[1,1,1]_{2(\frac{1}{2},\frac{1}{2})}
     +[2,0,0]_{(\frac{1}{2},\frac{1}{2})+(\frac{1}{2},\frac{3}{2})}
     $ \\ \hline $
     \dDelta+\frac{5}{2}$& $
     [0,0,1]_{(0,\frac{1}{2})+(0,\frac{3}{2})+2(1,\frac{1}{2})+(1,\frac{3}{2})+(2,\frac{1}{2})}
     +[0,0,3]_{(\frac{3}{2},0)}
     +[0,1,1]_{3(\frac{1}{2},0)+2(\frac{1}{2},1)+(\frac{3}{2},0)+(\frac{3}{2},1)}
     +[0,2,1]_{(0,\frac{1}{2})+(1,\frac{1}{2})}
     $\\& $
     +[1,0,0]_{(\frac{1}{2},0)+2(\frac{1}{2},1)+(\frac{1}{2},2)+(\frac{3}{2},0)+(\frac{3}{2},1)}
     +[1,0,2]_{(0,\frac{1}{2})+2(1,\frac{1}{2})}
     +[1,1,0]_{3(0,\frac{1}{2})+(0,\frac{3}{2})+2(1,\frac{1}{2})+(1,\frac{3}{2})}
     $\\& $
     +[1,1,2]_{(\frac{1}{2},0)}
     +[1,2,0]_{(\frac{1}{2},0)+(\frac{1}{2},1)}
     +[2,0,1]_{(\frac{1}{2},0)+2(\frac{1}{2},1)}
     +[2,1,1]_{(0,\frac{1}{2})}
     +[3,0,0]_{(0,\frac{3}{2})}
     $ \\ \hline $
     \dDelta+3$& $
     [0,0,0]_{(\frac{1}{2},\frac{1}{2})+(\frac{1}{2},\frac{3}{2})+(\frac{3}{2},\frac{1}{2})+(\frac{3}{2},\frac{3}{2})}
     +[0,0,2]_{2(0,0)+(1,0)+2(1,1)+(2,0)}
     +[0,1,0]_{3(0,1)+3(1,0)+2(1,1)+(1,2)+(2,1)}
     $\\& $
     +[0,1,2]_{2(\frac{1}{2},\frac{1}{2})+(\frac{3}{2},\frac{1}{2})}
     +[0,2,0]_{3(\frac{1}{2},\frac{1}{2})+(\frac{1}{2},\frac{3}{2})+(\frac{3}{2},\frac{1}{2})}
     +[0,2,2]_{(0,0)}
     +[0,3,0]_{(0,1)+(1,0)}
     $\\& $
     +[1,0,3]_{(1,0)}
     +[1,1,1]_{2(0,0)+2(0,1)+2(1,0)+2(1,1)}
     +[1,2,1]_{(\frac{1}{2},\frac{1}{2})}
     +[2,0,0]_{2(0,0)+(0,1)+(0,2)+2(1,1)}
     $\\& $
     +[2,0,2]_{(\frac{1}{2},\frac{1}{2})}
     +[2,1,0]_{2(\frac{1}{2},\frac{1}{2})+(\frac{1}{2},\frac{3}{2})}
     +[2,2,0]_{(0,0)}
     +[3,0,1]_{(0,1)}
     +[1,0,1]_{4(\frac{1}{2},\frac{1}{2})+2(\frac{1}{2},\frac{3}{2})+2(\frac{3}{2},\frac{1}{2})+(\frac{3}{2},\frac{3}{2})}
     $ \\ \hline $
     \dDelta+\frac{7}{2}$& $
     [0,0,1]_{2(\frac{1}{2},0)+3(\frac{1}{2},1)+(\frac{3}{2},0)+2(\frac{3}{2},1)+(\frac{3}{2},2)}
     +[0,0,3]_{(0,\frac{1}{2})+(1,\frac{1}{2})}
     +[0,1,1]_{3(0,\frac{1}{2})+(0,\frac{3}{2})+4(1,\frac{1}{2})+2(1,\frac{3}{2})+(2,\frac{1}{2})}
     $\\& $
     +[0,1,3]_{(\frac{1}{2},0)}
     +[0,2,1]_{2(\frac{1}{2},0)+2(\frac{1}{2},1)+(\frac{3}{2},0)}
     +[0,3,1]_{(0,\frac{1}{2})}
     +[1,0,0]_{2(0,\frac{1}{2})+(0,\frac{3}{2})+3(1,\frac{1}{2})+2(1,\frac{3}{2})+(2,\frac{3}{2})}
     $\\& $
     +[1,0,2]_{2(\frac{1}{2},0)+2(\frac{1}{2},1)+(\frac{3}{2},0)+(\frac{3}{2},1)}
     +[1,1,0]_{3(\frac{1}{2},0)+4(\frac{1}{2},1)+(\frac{1}{2},2)+(\frac{3}{2},0)+2(\frac{3}{2},1)}
     +[1,1,2]_{(0,\frac{1}{2})+(1,\frac{1}{2})}
     $\\& $
     +[1,2,0]_{2(0,\frac{1}{2})+(0,\frac{3}{2})+2(1,\frac{1}{2})}
     +[1,3,0]_{(\frac{1}{2},0)}
     +[2,0,1]_{2(0,\frac{1}{2})+(0,\frac{3}{2})+2(1,\frac{1}{2})+(1,\frac{3}{2})}
     +[2,1,1]_{(\frac{1}{2},0)+(\frac{1}{2},1)}
     $\\& $
     +[3,0,0]_{(\frac{1}{2},0)+(\frac{1}{2},1)}
     +[3,1,0]_{(0,\frac{1}{2})}
     $ \\ \hline $
     \dDelta+4$& $
     [0,0,0]_{3(0,0)+3(1,1)+(2,2)}
     +[0,0,2]_{3(\frac{1}{2},\frac{1}{2})+(\frac{1}{2},\frac{3}{2})+(\frac{3}{2},
     \frac{1}{2})+(\frac{3}{2},\frac{3}{2})}
     +[0,1,0]_{4(\frac{1}{2},\frac{1}{2})+2(\frac{1}{2},\frac{3}{2})+2(\frac{3}{2},
     \frac{1}{2})+2(\frac{3}{2},\frac{3}{2})}
     $\\& $
     +[0,1,2]_{(0,0)+2(0,1)+2(1,0)+(1,1)}
     +[0,2,0]_{3(0,0)+(0,1)+(0,2)+(1,0)+3(1,1)+(2,0)}
     +[0,2,2]_{(\frac{1}{2},\frac{1}{2})}
     $\\& $
     +[0,3,0]_{2(\frac{1}{2},\frac{1}{2})}
     +[0,4,0]_{(0,0)}
     +[1,0,1]_{(0,0)+3(0,1)+3(1,0)+4(1,1)+(1,2)+(2,1)}
     +[1,0,3]_{(\frac{1}{2},\frac{1}{2})}
     +[0,0,4]_{(0,0)}
     $\\& $
     +[1,1,1]_{4(\frac{1}{2},\frac{1}{2})+2(\frac{1}{2},\frac{3}{2})+2(\frac{3}{2},\frac{1}{2})}
     +[1,2,1]_{(0,0)+(0,1)+(1,0)}
     +[2,0,0]_{3(\frac{1}{2},\frac{1}{2})+(\frac{1}{2},
     \frac{3}{2})+(\frac{3}{2},\frac{1}{2})+(\frac{3}{2},\frac{3}{2})}
     $\\& $
     +[2,0,2]_{(0,0)+(1,1)}
     +[2,1,0]_{(0,0)+2(0,1)+2(1,0)+(1,1)}
     +[2,2,0]_{(\frac{1}{2},\frac{1}{2})}
     +[3,0,1]_{(\frac{1}{2},\frac{1}{2})}
     +[4,0,0]_{(0,0)}
     $ \\ \hline $
     \dDelta+\frac{9}{2}$& $
     [0,0,1]_{2(0,\frac{1}{2})+(0,\frac{3}{2})+3(1,\frac{1}{2})+2(1,\frac{3}{2})+(2,\frac{3}{2})}
     +[0,0,3]_{(\frac{1}{2},0)+(\frac{1}{2},1)}
     +[0,1,1]_{3(\frac{1}{2},0)+4(\frac{1}{2},1)+(\frac{1}{2},2)+(\frac{3}{2},0)+2(\frac{3}{2},1)}
     $\\& $
     +[0,1,3]_{(0,\frac{1}{2})}
     +[0,2,1]_{2(0,\frac{1}{2})+(0,\frac{3}{2})+2(1,\frac{1}{2})}
     +[0,3,1]_{(\frac{1}{2},0)}
     +[1,0,0]_{2(\frac{1}{2},0)+3(\frac{1}{2},1)+(\frac{3}{2},0)+2(\frac{3}{2},1)+(\frac{3}{2},2)}
     $\\& $
     +[1,0,2]_{2(0,\frac{1}{2})+(0,\frac{3}{2})+2(1,\frac{1}{2})+(1,\frac{3}{2})}
     +[1,1,0]_{3(0,\frac{1}{2})+(0,\frac{3}{2})+4(1,\frac{1}{2})+2(1,\frac{3}{2})+(2,\frac{1}{2})}
     +[1,1,2]_{(\frac{1}{2},0)+(\frac{1}{2},1)}
     $\\& $
     +[1,2,0]_{2(\frac{1}{2},0)+2(\frac{1}{2},1)+(\frac{3}{2},0)}
     +[1,3,0]_{(0,\frac{1}{2})}
     +[2,0,1]_{2(\frac{1}{2},0)+2(\frac{1}{2},1)+(\frac{3}{2},0)+(\frac{3}{2},1)}
     +[2,1,1]_{(0,\frac{1}{2})+(1,\frac{1}{2})}
     $\\& $
     +[3,0,0]_{(0,\frac{1}{2})+(1,\frac{1}{2})}
     +[3,1,0]_{(\frac{1}{2},0)}
     $ \\ \hline $
     \dDelta+5$& $
     [0,0,0]_{(\frac{1}{2},\frac{1}{2})+(\frac{1}{2},\frac{3}{2})+(\frac{3}{2},\frac{1}{2})+(\frac{3}{2},\frac{3}{2})}
     +[0,0,2]_{2(0,0)+(0,1)+(0,2)+2(1,1)}
     +[0,1,0]_{3(0,1)+3(1,0)+2(1,1)+(1,2)+(2,1)}
     $\\& $
     +[0,1,2]_{2(\frac{1}{2},\frac{1}{2})+(\frac{1}{2},\frac{3}{2})}
     +[0,2,0]_{3(\frac{1}{2},\frac{1}{2})+(\frac{1}{2},\frac{3}{2})+(\frac{3}{2},\frac{1}{2})}
     +[0,2,2]_{(0,0)}
     +[0,3,0]_{(0,1)+(1,0)}
     $\\& $
     +[1,0,3]_{(0,1)}
     +[1,1,1]_{2(0,0)+2(0,1)+2(1,0)+2(1,1)}
     +[1,2,1]_{(\frac{1}{2},\frac{1}{2})}
     +[2,0,0]_{2(0,0)+(1,0)+2(1,1)+(2,0)}
     $\\& $
     +[2,0,2]_{(\frac{1}{2},\frac{1}{2})}
     +[2,1,0]_{2(\frac{1}{2},\frac{1}{2})+(\frac{3}{2},\frac{1}{2})}
     +[2,2,0]_{(0,0)}
     +[3,0,1]_{(1,0)}
     +[1,0,1]_{4(\frac{1}{2},\frac{1}{2})+2(\frac{1}{2},\frac{3}{2})+2(\frac{3}{2},\frac{1}{2})+(\frac{3}{2},\frac{3}{2})}
     $ \\ \hline $
     \dDelta+\frac{11}{2}$& $
     [0,0,1]_{(\frac{1}{2},0)+2(\frac{1}{2},1)+(\frac{1}{2},2)+(\frac{3}{2},0)+(\frac{3}{2},1)}
     +[0,0,3]_{(0,\frac{3}{2})}
     +[0,1,1]_{3(0,\frac{1}{2})+(0,\frac{3}{2})+2(1,\frac{1}{2})+(1,\frac{3}{2})}
     +[0,2,1]_{(\frac{1}{2},0)+(\frac{1}{2},1)}
     $\\& $
     +[1,0,0]_{(0,\frac{1}{2})+(0,\frac{3}{2})+2(1,\frac{1}{2})+(1,\frac{3}{2})+(2,\frac{1}{2})}
     +[1,0,2]_{(\frac{1}{2},0)+2(\frac{1}{2},1)}
     +[1,1,0]_{3(\frac{1}{2},0)+2(\frac{1}{2},1)+(\frac{3}{2},0)+(\frac{3}{2},1)}
     $\\& $
     +[1,1,2]_{(0,\frac{1}{2})}
     +[1,2,0]_{(0,\frac{1}{2})+(1,\frac{1}{2})}
     +[2,0,1]_{(0,\frac{1}{2})+2(1,\frac{1}{2})}
     +[2,1,1]_{(\frac{1}{2},0)}
     +[3,0,0]_{(\frac{3}{2},0)}
     $ \\ \hline $
     \dDelta+6$& $
     [0,0,0]_{(0,0)+(0,2)+(1,1)+(2,0)}
     +[0,0,2]_{(\frac{1}{2},\frac{1}{2})+(\frac{1}{2},\frac{3}{2})}
     +[0,1,0]_{2(\frac{1}{2},\frac{1}{2})+(\frac{1}{2},\frac{3}{2})+(\frac{3}{2},\frac{1}{2})}    +[2,0,2]_{(0,0)}
     +[2,1,0]_{(1,0)}
     $\\& $
     +[0,1,2]_{(0,1)}
     +[0,2,0]_{2(0,0)+(1,1)}
     +[1,0,1]_{(0,0)+2(0,1)+2(1,0)+(1,1)}
     +[1,1,1]_{2(\frac{1}{2},\frac{1}{2})}
     +[2,0,0]_{(\frac{1}{2},\frac{1}{2})+(\frac{3}{2},\frac{1}{2})}
     $ \\ \hline $
     \dDelta+\frac{13}{2}$& $
     [0,0,1]_{(0,\frac{1}{2})+(0,\frac{3}{2})+(1,\frac{1}{2})}
     +[0,1,1]_{(\frac{1}{2},0)+(\frac{1}{2},1)}
     +[1,0,0]_{(\frac{1}{2},0)+(\frac{1}{2},1)+(\frac{3}{2},0)}
     +[1,0,2]_{(0,\frac{1}{2})}
     $\\& $
     +[1,1,0]_{(0,\frac{1}{2})+(1,\frac{1}{2})}
     +[2,0,1]_{(\frac{1}{2},0)}
     $ \\ \hline $
     \dDelta+7$& $
     [0,0,0]_{(\frac{1}{2},\frac{1}{2})}
     +[0,0,2]_{(0,0)}
     +[0,1,0]_{(0,1)+(1,0)}
     +[1,0,1]_{(\frac{1}{2},\frac{1}{2})}
     +[2,0,0]_{(0,0)}
     $ \\ \hline $
     \dDelta+\frac{15}{2}$ & $
     [0,0,1]_{(\frac{1}{2},0)}
     +[1,0,0]_{(0,\frac{1}{2})}
     $ \\ \hline $
     \dDelta+8$& $
     [0,0,0]_{(0,0)}
     $ \\ \hline
     \end{tabular}}
 \caption{Long Konishi multiplet \label{Ktable}}
\end{table}

%%%%%%%%%%%%%%%%%%%%%%%%%%%%%%%%%%%%%%%%%%%%%%%%%%%%%%%%

\bigskip
%%%%%%%%%%%%%%%%%%%%%%%%%%%%%%%

\end{document}